\documentclass[12pt, a4paper]{article}
\usepackage{graphicx}
\usepackage{mathrsfs}
\usepackage{amssymb}
\usepackage{color}
\usepackage{amsmath}
\usepackage{ascmac}
\usepackage{amsthm}
\usepackage{booktabs}
\usepackage{lscape}
\usepackage{dcolumn}
\usepackage{longtable}
\usepackage{dcolumn}
\usepackage{arydshln}
\usepackage{bm}
\usepackage{url}
\usepackage{caption}
\usepackage{subfig}
\usepackage{setspace}
\usepackage{cases}
\usepackage{comment}
\usepackage{bigfoot}

\newcommand{\vx}{\mathbf{x}}

\newcommand{\eps}{\epsilon}

\newcommand{\vgamma}{\bm{\gamma}}

\begin{document}

\title{\LARGE{Consumption Smoothing in the Working-Class Households of Interwar Japan}}

\author{Kota Ogasawara\thanks{Department of Industrial Engineering and Economics, School of Engineering, Tokyo Institute of Technology, 2-12-1, Ookayama, Meguro-ku, Tokyo 152-8552, Japan (E-mail: ogasawara.k.ab@m.titech.ac.jp).\newline
I wish to thank Dan Bogart, Bishnupriya Gupta, Peter Lindert, Kazushige Matsuda, Stefan \"Oberg, Tetsuji Okazaki, Sakari Saaritsa, Eric Schneider, Masayuki Tanimoto, Ken Yamada, Elise van Nederveen Meerkerk, Hans-Joachim Voth,  and the four anonymous referees.~I would also like to thank the seminar participants at the University of Tokyo; Montr\'eal; Massachusetts Institute of Technology; and Paris School of Economics for their helpful comments.
I thank Ryo Nagaya for research assistance.
There are no conflicts of interest to declare.
All errors are my own.}}
\date{\today}
\maketitle

\begin{abstract}
\begin{spacing}{0.85}
I analyze factory worker households in the early 1920s in Osaka to examine idiosyncratic income shocks and consumption.
Using the household-level monthly panel dataset, I find that while households could not fully cope with idiosyncratic income shocks at that time, they mitigated fluctuations in indispensable consumption during economic hardship.
In terms of risk-coping mechanisms, I find suggestive evidence that savings institutions helped mitigate vulnerabilities and that both using borrowing institutions and adjusting labor supply served as risk-coping strategies among households with less savings.
\end{spacing}
\bigskip

\noindent\textbf{Keywords:}
consumption smoothing;
precautionary saving;
risk-coping strategies;
risk-sharing
\bigskip

\noindent\textbf{JEL Codes:}
E21; 
N35 

\end{abstract}

\newpage
\section{Introduction}

The insurance contracts offered through market and nonmarket mechanisms are important strategies for households to smooth their consumption in the face of idiosyncratic shocks.
A collapse of consumption smoothing impedes human capital accumulation and the demographic structure itself, especially in developing economies (Foster 1995; Rose 1999; Gertler and Gruber 2002; Dercon and Krishnan 2002).
Therefore, households' risk-coping behavior has been widely studied, ever since a series of pioneering studies provided a systematic empirical design to test consumption smoothing (Rosenzweig 1988; Mace 1991; Cochrane 1991).\footnote{For instance, Townsend (1994) found that risk sharing among three villages in southern India insured household-level consumption against idiosyncratic shocks. Fafchamps et al. (1998) and Fafchamps and Lund (2003) showed that livestock transactions, gifts, and loans from informal networks are used to mitigate idiosyncratic shocks among rural households in Burkina Faso and the Philippines. As for urban households, Skoufias and Quisumbing (2005) provided evidence that both borrowing and additional labor supply are used for self-insurance in Russia. See Townsend (1995) and Dercon (2004) for reviews. Attanasio and Pistaferri (2016) and Meyer and Sullivan (2017) provide the latest discussions on consumption and income inequality.}

In the past, risk-sharing institutions and self-insurance behavior also played an important role in allowing working-class households to cope with idiosyncratic shocks during periods of economic development.\footnote{Kiesling (1996) determined the importance of informal sources of income assistance (e.g., savings, transfers, and charity) in Victorian Lancashire (see also Boyer (1997) for an alternative view). Horrell and Oxley (2000) found evidence that British industrial households used supportive organizations, such as sickness and health benefit clubs, in the late 19th century. Scott and Walker (2012) also found that interwar British working-class households frequently used risk-sharing institutions, such as clubs and hire purchases, to smooth expenditures.}
The study systematically investigates the risk-coping behavior of working-class urban households.
To do so, I digitize a detailed monthly longitudinal budget survey on factory worker households conducted after World War I in Osaka, the second largest city in Japan.\footnote{According to the first Population Census of 1920, Tokyo and Osaka had $2,173,201$ and $1,252,983$ citizens, respectively (Statistics Bureau of the Cabinet 1925, 1929c). These two cities accounted for approximately $34$\% of Japan's urban population at the time (Statistics Bureau of the Cabinet 1929a).}
Utilizing an empirical design that exploits the within variations in the households' consumption and income, I estimate the income elasticity to determine the extent of consumption insurance.
The result indicates that the factory worker households could not fully deal with idiosyncratic shocks.
Their income elasticity of total consumption expenditure is found to be comparable to that of urban households in developing countries at the same development stage.
Nonetheless, the income elasticities of consumption by subcategory suggest that they might have mitigated the fluctuations in indispensable consumption to some degree.
While the elasticities for luxury categories such as furniture, clothes, and entertainment expenses were greater, payments for a few indispensable categories such as food and housing were clearly inelastic.
I also provide evidence suggesting that savings institutions helped mitigate idiosyncratic shocks.
The households precautionarily saved the surplus built up and relied more on withdrawals from savings than borrowing when they faced the shocks.
Temporary income from borrowing, particularly from pawnshops, as well as from the additional labor supply of the wife and child had also played a role among relatively vulnerable households.

This current study contributes to the literature in the following two ways.
First, it contributes to the historical literature by adopting a systematic empirical design to test risk-sharing, which can be used in future economic history studies.
Different types of analytical specifications have thus far been applied in the field of economic history, which makes comparing the degrees of risk sharing in different countries and eras complicated.
In the current study, I employ a standardized empirical design to test risk-sharing (Cochrane 1991) and compare the estimated income elasticities with those in modern developing societies.\footnote{Regarding consumption smoothing internationally over time, Persaud (2019), the most recent study on this topic, revealed that Indian indentureship contracts for South-South migration are used as a device to mitigate the risk of volatility in economic outcomes in India. However, while Persaud (2019) focused on coping with the risks of macroeconomic shocks, this study investigates consumption smoothing in the face of idiosyncratic shocks.}
The degree of consumption smoothing can reflect the maturity of insurance markets in a given economy (Dercon 2004).
Therefore, applying this approach to other historical panel datasets will offer comparable estimates for a diverse range of economies.

Second, the current study uses a household-level monthly expenditure panel dataset.
Since the concept of consumption smoothing pertains to the dynamics of household behavior over time, panel data on household budgets are preferable to test households' smoothing behavior (Mace 1991).
However, most previous economic history studies have used cross-sectional data rather than panel data.\footnote{Owing to data unavailability, it was difficult to compile the household-level longitudinal budget data in the historical context (see James and Suto (2011) and Scott and Walker (2012) for discussions on this). For example, Nakagawa (1985), who investigated consumption patterns in urban factory worker households in the early 20th century in Japan, compared average expenditure on certain consumption categories documented in several cross-sectional survey reports measured in different years. While an exceptional study by Saaritsa (2011) used a quarterly panel dataset of 142 households from Helsinki in 1928, the current study uses data aggregated at a higher frequency to identify households' short-run responses.}
To bridge this gap, I use monthly variations in the household budget to investigate risk-coping behavior among working-class households.
Relatedly, Japan is a suitable country for investigating risk-coping behavior for the following reasons.
While national health insurance schemes expanded in many European countries between the end of the 19th century and early 20th century, Japan did not establish such a scheme until the mid-20th century (Bowblis 2010).
In addition, while the prevailing fragile labor contracts resulted in high turnover rates among factory workers, the unemployment insurance bill was not passed until the late 20th century. 
Therefore, idiosyncratic income shocks cannot be compensated by public insurance schemes (Section~\ref{sec:sec21}).
This feature of prewar Japan offers an ideal environment for studying consumption-smoothing responses to idiosyncratic income shocks.

The remainder of this paper is organized as follows. 
Section~\ref{sec:sec2} briefly reviews the historical context and explains the risk-coping devices.
Section~\ref{sec:sec3} introduces the data used and discusses the sample characteristics.
Section~\ref{sec:sec4} empirically analyzes consumption smoothing and Section~\ref{sec:sec5} assesses risk-coping mechanisms.
Section~\ref{sec:sec6} concludes.

\section{Background} \label{sec:sec2}

\subsection{The Nature of Risk} \label{sec:sec21}

In early 20th-century Japan, employment contracts were typically fragile and stipulated no fixed term of employment; thus, the labor mobility of factory workers, comprising not only unskilled workers but also skilled workers, was extremely high (Moriguchi 2000).
After World War I, large companies began to introduce comprehensive corporate welfare programs for factory workers to accumulate firm-specific human capital, and the Retirement Allowance Fund Law of 1936, which obligated employers to set up a retirement allowance fund for their employees, complemented these enterprise-based welfare programs (Moriguchi 2003).
As a result, the average annual turnover rates of large companies began to decline after the war and fell to below 10\% in the late 1920s (Hyodo 1971).

However, the factory workers employed by such large companies comprised only about 20\% of all production workers in the late 1920s (Moriguchi 2003, p.~644).
Additionally, the average annual turnover rate of small and medium-sized enterprises was still approximately 30\% (Hyodo 1971; Taira 1970).
Therefore, except for a few favored workers among large companies, worker uncertainty in the labor market was high in interwar Japan, a period in which no comprehensive social security system had yet been established (Odaka 1999).\footnote{In general, the introduction of government social insurance schemes reduced the purchase of private insurance and the use of precautionary savings (Kantor and Fishback 1996; Emery 2010). In interwar London, for example, pension payments constituted an income below the poverty line (Baines and Johnson 1999). However, comprehensive public assistance did not exist in prewar Japan (Kase 2006); for example, the unemployment insurance bill was not passed until 1947. This underscores the importance of risk-sharing institutions at that time.}

In addition to unemployment, sickness and injury could be another source of uncertainty for workers.
The Factory Act of 1916 included an article setting an allowance for workers taking medical leave.
In reality, however, most of the factories subject to the Act did not fully compensate workers who took medical leave.\footnote{For example, many of the factories in Osaka that were subject to the Act paid workers only half of their daily wages for medical leave of up to three months (Department of Social Affairs, Osaka City Office 1923~p.~203).}
Moreover, because the Act applied only to factories with 15 or more workers, workers in many small factories were entirely neglected.\footnote{The former Health Insurance Act was enacted in 1927, which extended the scope of the sicknesses and injuries covered by the Factory Act. However, it was still designed only for workers at factories subject to the Factory Act.}
Thus, sicknesses and injuries might have caused a certain reduction in factory workers' incomes.

Despite these uncertainties, Japan's economic growth remained stable during the interwar period (Nakamura 1981).
Inequality among members of the working class, measured in terms of the Gini coefficient, decreased between the early 1920s and early 1930s (Yazawa 2004; Bassino 2006).
Consequently, representative measures of human and physical capital accumulation such as average years of education and children's height climbed steadily during this period.\footnote{See Godo (2013) and Schneider and Ogasawara (2018), respectively. This accumulation of human and physical capital became a driving force of economic growth in postwar Japan (World Bank 1993).}
These developmental outcomes suggest that working-class households might have coped with the risks to a certain extent.

\subsection{Risk-Coping Devices} \label{sec:sec22}

\subsubsection*{\textit{Savings Institutions}}

Precautionary savings are an important primary risk-coping device (Deaton 1991; Carroll 1997; Carroll et al. 2003).
In prewar Japan, postal savings (\textit{y\=ubin chokin}) and savings banks (\textit{chochiku gink\=o}) were widely used saving institutions around 1920 (Okazaki 2002; Tanaka 2012).
In Osaka city, there were $117$ postal savings offices and $55$ savings banks in 1920.
The number of people with postal savings and savings bank accounts were reported to be $725,642$ and $1,648,150$, respectively, accounting for approximately $60$\% and $130$\% of the number of citizens measured in the census.\footnote{See Osaka City Office (1922, pp.~6(59); 6(61); 6(67)) and Statistical Bureau of the Cabinet (1925, p.~2).}
Although the statistics must double count people having multiple accounts, they suggest that a large proportion of workers had savings accounts.
Average savings per manufacturing worker were roughly $50$ yen and $35$ yen in postal savings and savings banks in 1920, respectively, approximately equivalent to $30$-$50$\% of the average monthly income of urban factory workers.\footnote{See Osaka City Office (1922, pp.~6(63); 6(67)), Bureau of Social Welfare (1923, pp.~24--39), and Postal Savings Bureau (1924, p.~72) for the statistics.}
This implies that, despite having only small amounts of savings, factory workers might have drawn on their savings in the event of economic hardship, as suggested by James and Suto (2011).\footnote{They found that savings correlated positively with transitory income in cross-sectional household surveys from the 1910s and 1920s in Japan. According to their estimates, the savings level was higher than that of working-class households in the United States in the early 20th century. They suggested that the widespread use of postal savings accounts in prewar Japan encouraged or facilitated saving relative to the institutions available to US workers.}

Another type of savings institution was mutual loan associations (\textit{mujin}), in which members deposited a fixed amount of money into a unit and withdrew it according to certain association rules.
However, the association distributed deposits to members using either lotteries or bidding.
This means that members could not withdraw the right deposits at the right time.
In addition, a large deposit per unit (usually $100$--$300$ yen) meant most members of \textit{mujin} were probably owners of small businesses, such as merchants.\footnote{See Osaka City Office (1934, p.~250) and Banks Division, Secretariat of Finance Ministry (1915, pp.~64--69; 246).}
Moreover, less than 2\% of the working population in Osaka in 1915 invested money in this way (Osaka City Office 1934, p.~251).
In this light, unlike postal savings and savings banks, mutual loan associations were not useful risk-coping devices for factory workers, a large and growing proportion of the urban population.

\subsubsection*{\textit{Lending Institutions}}

Urban working-class households in prewar Japan could access a few lending institutions.\footnote{I examine the lending institutions in cities. See Saito (1989) and Shibuya (2000) for the local contexts of these institutions.}
Among these, pawnshops (\textit{shichiya}) were the most popular lending institutions, especially among factory workers.\footnote{While pawnshops had been in decline in the early 20th-century Europe, they remained a popular lending institution throughout 20th-century Japan (Murhem 2015; Shibuya et al.~1982). I focus on private pawnshops because public pawnshops only officially began to open in the late 1920s when the Public Pawnshop Law was enacted in 1927. See Shibuya et al.~(1982, pp. 412--446) for the historical context of public pawnshops.}
Shibuya et al.~(1982) argued that factory workers were the most frequent users of pawnshops in cities for a few main reasons (pp.~328--338).
First, pawnshops were easier to access than other lending institutions because lenders did not need to check the credit of borrowers.
Furthermore, since inexpensive clothes were the most common pawns at that time, workers could borrow money without the risk of falling into heavy debt.
Second, the interest rates at pawnshops were relatively low as they were regulated by the Pawnbroker Regulation Act of 1895.
Accordingly, the redemption rate was substantially high: approximately nine out of 10 borrowers could repay their loans within the short term, which enabled them to minimize their interest payments.
Finally, physical accessibility was sufficiently high because the number of pawnshops was substantially greater than that of other lending institutions such as mutual loan associations and usuries.\footnote{Inoue (2021, p.~5) revealed that the average numbers of pawnshops, ordinary banks, mutual loan institutions, and post offices in each prefecture in 1924 were $379$, $148$, $6$, and $184$, respectively.}

In Osaka city, pawnshops were regarded as a major lending institution among factory workers (Osaka City Office 1920, p.~153).
Of the 983 pawn shops in October 1919, factory workers accounted for $60$\% of all users and $43$\% of the borrowing.\footnote{See Osaka City Office (1920, pp.~102; 104; 122; 123; 136; 138) for the statistics used in this paragraph.}
More than 80\% of those loans were from the clothes and the redemption rate was approximately 95\%.\footnote{Shibuya et al., (1982,~pp. 335--338), analyzing the ledgers of a pawnshop in Osaka, also showed that more than 90\% of all articles for pawning were clothes.
Given the long-term value of a kimono, for example, networks among specialized dealers, sellers, and secondhand clothes shops sprang up, which eventually formed a large secondhand market in prewar Japan (Francks 2012, p.~161). Given this market, pawnshops could evaluate the pawned items against prices in the secondhand market (Bank of Japan 1913).}
The average amount borrowed by factory workers per event was $11$ yen, accounting for roughly $10$\% of the mean monthly income of factory workers at that time.
These amounts suggest that those households used pawnshops to smooth their consumption when they faced idiosyncratic shocks rather than to buy luxury goods or invest, which would require much higher borrowing.

Money lenders (\textit{kinsen kashitsuke gy\=o}) also existed in this period (Osaka City Office 1934, pp. 182--183).
In 1926, there were 196 money lenders, but the number of lending events per lender was only $118$, substantially smaller than that of pawnshops (i.e., approximately $4,000$ pawns per shop in the same year).
Furthermore, while the average loan amount per pawn was usually $5$--$10$ yen in pawnshops, this figure was substantially large for money lenders at slightly under $300$ yen, approximately three times the average monthly household income of factory workers.\footnote{See Osaka City Office (1934, pp.~203; 241) and Osaka City Office (1927, p.~6(85)) for the statistics.}
This means that money lenders were predominantly used by business owners, such as the owners of small enterprises, merchants, and landowners (Shibuya 2000, pp.~184; 248).
Indeed, by the early 1920s, most money lenders had become usuries (\textit{k\=origashi}).\footnote{Usury interest rates were considerably higher than those of pawnshops. For instance, if the loan amount was $10$ yen, the average annual interest rate was around 30\% in pawnshops compared with more than 100\% in usuries (Shibuya et al.~1982, p.~348; Shibuya 2000, pp.~606--607). Given their astronomical interest rates, usuries were regarded as risky lending institutions for working-class households. The Social Welfare Department of Osaka stated that ``usuries are literary `vampires', and eradicating them is an urgent issue in social policy'' (Osaka City Office 1926, p.~60).}
Since usuries expected to siphon money from those guarantors rather than borrowers themselves, three or more joint guarantors with credibility were required to access those lenders.
This feature made it harder for the working class to use usuries.

Finally, the main customers of ordinary banks and credit unions (\textit{shiny\=o kumiai}) at that time were companies, including other banks.
In 1920, there were 24 ordinary banks, with an average loan amount of approximately $27$ thousand yen; further, a large part of their collateral was stock certificates.\footnote{See Osaka City Office (1922, pp.~6(28); 6(34); 6(35)) for the statistics.}
Although the 24 credit unions might have been accessible to workers, their $4,451$ members accounted for only $0.4$\% of the working population in Osaka city in 1926 (Osaka City Office 1934, p. 276).

\subsubsection*{\textit{Informal Insurance Provided by Networks}}

Evidence on rural economies in today's developing countries has highlighted the importance of informal networks (e.g., gifts from friends and family) for coping with idiosyncratic shocks because formal insurance markets rarely exist in village economies (Rosenzweig 1988).\footnote{See also Munshi and Rosenzweig (2016) and Attanasio and Krutikova (2020) for recent work on the role of informal networks in village economies in the developing world.}
Systematic statistics on gift from personal networks in prewar Japan are unavailable.
However, a survey of $185$ factory worker households in Tokyo city and surrounding suburban municipalities in November 1922 found that the average monthly income from personal networks was $2.9$ yen, accounting for $2.8$\% of total monthly income (Social Affairs Division 1925, pp.~58--59).\footnote{Tokyo prefecture asked wide variety of public offices, companies, schools, banks, and factories to sample working-class households with 2--8 family members and earning 60--250 yen per month. While both conditions were not necessarily met, $1,027$ households (including $185$ factory worker households) were sampled (Social Affairs Division 1925, pp.~3--16). Despite the cross-sectional survey being conducted in a specific month, November is a useful month to study the income structure because there was no specific seasonality in that month in prewar Japan (Section~\ref{sec:sec33}). In addition, average monthly income ($104$ yen) in the sampled factory worker households was in a similar range to those of the other representative surveys (Panel B in Online Appendix Table~\ref{tab:taba2}).}
This suggests that using temporary income from personal networks might have been a useful strategy for urban factory worker households.

\subsubsection*{\textit{Labor Supply Adjustments}}

The traditional and still prevalent view of the labor supply among urban working-class households in prewar Japan is the male breadwinner model.
In that period, the wife's elasticity of labor supply was considerably low and children rarely worked compared with the extensive use of child labor during the Industrial Revolution in Europe (Saito 1995; 1996; Chimoto 2012).
However, a cross-sectional analysis of poor households in Tokyo found a negative correlation between the income of the household head and working status of children aged 15 years and older (Yazawa 2004, p.~315).
Although this is a suggestive result, one must be careful about the fact that all existing studies have relied on cross-sectional data, meaning that the dynamic compensating responses to idiosyncratic shocks to the household head's earnings have not yet been investigated.
Considering this, I quantify the labor supply of the wife and children in response to shocks to the income of the household head in Section~\ref{sec:sec52}.

To summarize, working-class households were able to use formal and informal risk-coping strategies.
Although informal insurance via private networks is not documented as a separate income category, I consider three categories (i.e., withdrawals and gifts, borrowing, and labor supply adjustments) when I evaluate those strategies in Section~\ref{sec:sec5}. I do so using the best available household-level longitudinal budget dataset, which I will introduce in the following section.

\section{Data}\label{sec:sec3}

\subsection{The Sample}\label{sec:sec31}

I compiled the data on the monthly budget of working-class households between July 1919 and July 1920 from the Report of Labor Research (RLR).\footnote{I use a reprinted edition of the original archives, as found in Tada (1991a). Although James and Suto (2011) employed 99 households from the RLR as an annual cross-sectional dataset, the current study is the first to digitize monthly panel data from this survey.}
In this survey, the Municipal Bureau of Labor Research of Osaka investigated the monthly income and expenditure of 411 wage-earning and salaried workers in the Osaka city area.
Although the details of the sampling method were not recorded, as is the case with other historical household survey datasets, households were selected through a labor union or directly at factories.
To investigate the features of the sample, I compare the occupations among household heads within RLR households with statistics obtained from the national population census.
As shown in column (1) of panel A in Table~\ref{tab:tab1}, only $14$\% of men in the workforce across Osaka prefecture worked in the agricultural sector compared with the national figure of $46$\%.
Column (2) also indicates that 90\% of working men in Osaka city worked in the manufacturing, commerce, and transportation industries, compared with the national figure of 40\%.\footnote{These national figures are obtained from the Statistics Bureau of the Cabinet~(1929a,~pp.~8--11; 1929b,~pp.~8--11).}
This reflects that Osaka was an industrialized city and a nice setting to study the consumption decisions of industrial workers.\footnote{Similar data are not available for Tokyo, a city as industrialized as Osaka at that time, to my knowledge.}
In the RLR dataset, information on the household head's occupation can be obtained for $406$ of the $411$ households.
Column (3) shows that these heads mainly worked in the manufacturing industry; $83$\% were factory workers.

Considering this feature of sampling, the current study targets only factory worker households.
I first extracted households whose heads worked in factories during the initial survey period.\footnote{
I focused on households classified as involved in the manufacturing industry throughout the sample period to analyze the dynamic behavior of factory worker households. Among the 237 households used in the analysis, there were only six heads who changed their occupation within one month of the sample period. However, these temporary changes only slightly affected my main results (see Online Appendix~\ref{sec:secb1} for details).}
Correspondingly, 335 of the original 411 households remain.
I then excluded the 18 households lacking information on monthly income or family structure.
Further, 78 households were dropped because they were observed for only one month during the survey period.
Finally, two households were dropped on account of having reported unrealistic income and expenditure values (i.e., exceeding 500 yen per month).
Accordingly, data on 237 factory worker households were used in the empirical analyses.\footnote{I conducted a two-sample $t$-test with unequal variances between the 237 households and the rest of the original 411 households. The differences in the total income, expenditure, and family size control variables (Panel C of Table~\ref{tab:sum}) are not statistically significant at the 5\% level. Although I focus on factory worker households, this finding implies that RLR households have similar characteristics and thus can be classified into a similar social class.}

The mean household size of the RLR sample is $4.00$, whereas that of the factory worker households in the manufacturing industry of Osaka city taken from the population census is $3.99$.
Hence, my sample does not contain outliers or otherwise unusual household size values.
Column (1) in panel B of Table~\ref{tab:tab1} shows the percentage share of adult male factory workers in each manufacturing sector in Osaka city in 1920 and column (3) lists the percentage share among RLR household heads.
The heads are more likely to work in the machine sector than in other sectors, consistent with the descriptions in the original RLR document.
Column (2) lists the average monthly income of adult male factory workers,\footnote{These figures are calculated based on the wage statistics of the manufacturing census. See Online Appendix~\ref{sec:seca2} for the finer details of the calculation.} whereas column (4) lists the average monthly income of the household heads in the RLR sample.
Clearly, the average monthly income of adult males is close to the monthly income of the RLR heads in each sector.\footnote{The textile and chemical sectors have a slightly larger difference because seven households report a higher income of the head---more than $150$ yen per month. However, all analytical results are materially similar if I exclude these households. Thus, these differences in the textile and chemical sectors do not hinder my main findings.}
In addition, the average monthly incomes of adult males and the monthly incomes of the heads show similar figures across different sectors.
Both support the evidence that the disproportion in the occupation structure in the RLR sample does not lead to an upward income bias and that the RLR sample approximates the average factory worker in the manufacturing sector in Osaka city.\footnote{
In Online Appendix~\ref{sec:seca3}, I provide additional evidence that the monthly household income and expenditure, and household size of the RLR sample, are close to those of the average factory worker households in the representative large cities.}
Even if there are some differences with other factory workers, or with the laboring population remain, the RLR sample provides a rare opportunity to study consumption and risk coping strategies in an industrializing economy.

\subsection{Variables}\label{sec:sec32}

The key data used in this study are monthly household consumption and income.
The RLR reports 10 consumption subcategories: food, housing, utilities, furniture, clothes, education, medical, entertainment, transportation, and miscellaneous.
These consumption variables are mainly used to estimate the income elasticities in Section~\ref{sec:sec4}.
Other expenditures are divided into tax payments, liquidation of loans, and deposits to savings.
On the contrary, income is divided into six categories: head, wife, child, other,\footnote{The ``other'' category includes the income of other family members (i.e., other than the head, wife, and children) and income from boarding and lodging. However, this subcategory is negligible because it is zero in almost all of the households.} borrowing, and miscellaneous.
The borrowing category includes money from lending institutions, while the miscellaneous category includes money from savings withdrawals and gifts (Municipal Bureau of Labor Research of Osaka 1921, pp. 22--27).
Therefore, I rename the miscellaneous category, calling it withdrawals and gifts (not divisible).
These income variables are used in the mechanism analysis in Section~\ref{sec:sec5}.

The quality of the data is considered to be sufficiently high to conduct quantitative analyses.
The investigators visited all households and instructed them all once or twice per month to check the account books and maintain the quality of the survey (Tada 1991a, pp. 11--12).
Hence, lazy respondents were unlikely to simply copy and paste entries rather than record each purchase.
If this sort of repetition did occur, however, the reported expenses would take the same values during the survey period.
To test this potential issue, I check whether the first-differenced values of food expenses have sufficient variation for all the households.
The food category is the most useful category for this exercise because it requires careful bookkeeping to sort many grocery items; further, expenditure on food has the largest share of the total as well as large variations due to seasonality and price changes.\footnote{Categories that have fixed and/or censoring natures are unsuitable for such a test. For example, housing expenses rarely change because rent is likely to be fixed. Similarly, expenses on education and transportation tend to take zero values in some households because they should not be unobserved in households without children and in those commuting by walking, respectively.}
I find that the observations of three households take the same values across two consecutive months, accounting for only $0.16$\% of all observations.
This supports the evidence that measures of consumption are less likely to contain measurement errors.

\subsection{Trends}\label{sec:sec33}

\begin{figure}[]
\centering
\subfloat[Income and expenditure (in yen)]{\label{fig:mincexp}\includegraphics[width=0.55\textwidth]{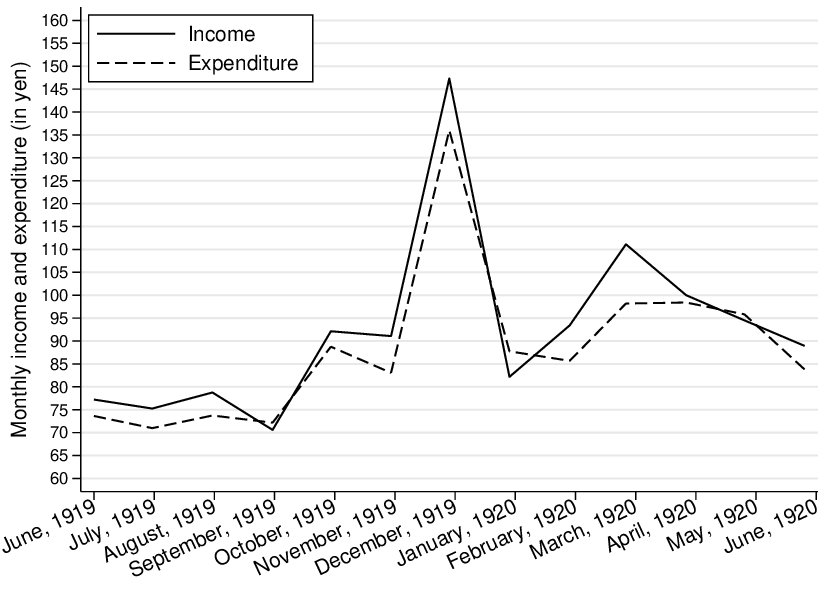}}\\
\subfloat[Income minus expenditure (in yen)]{\label{fig:mdeff}\includegraphics[width=0.55\textwidth]{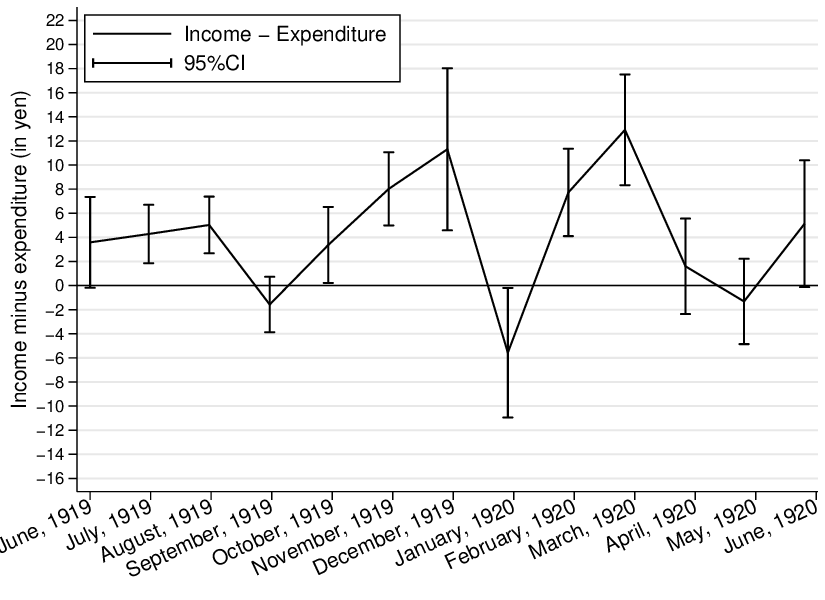}}
\caption{Monthly income and expenditure \\between June 1919 and June 1920}
\label{fig:monthly}
\scriptsize{\begin{minipage}{350pt}
Notes: Monthly income and expenditure are illustrated in Figure~\ref{fig:mincexp}.
The difference between monthly income and monthly expenditure is shown in Figure~\ref{fig:mdeff}.
Sources: Created by the author from the Municipal Bureau of Labor Research of Osaka (1919--1920).
\end{minipage}}
\end{figure}

Figure~\ref{fig:mincexp} illustrates the rising monthly income and expenditure during the sample period.
These trends are consistent with the increasing living standards after World War I (Nakamura and Odaka 1989, pp. 36--37).
One may wonder whether this upward trend in income suggests that households had more savings or less debt and thus were better able to deal with shocks.
However, moderate inflation during the sample period partially canceled out the benefit of the upward trend.\footnote{The Bank of Japan (1986, pp.~436--438) suggests that prices increased by approximately 4\% from 1919 to 1920, whereas the increase in aggregate household expenditure of RLR households was approximately 6\% in the same period.}
Moreover, Figure~\ref{fig:mincexp} indicates that both income and expenditure decreased after April 1920.
This trend reflects the recession that followed the war after March 1920 in Japan (Takeda 2002, pp.~9--11).
Therefore, my sample period includes expansions and recessions, which should offer an ideal setting within which to analyze the risk-coping behavior of households.

\begin{figure}[]
\centering
\includegraphics[width=10cm]{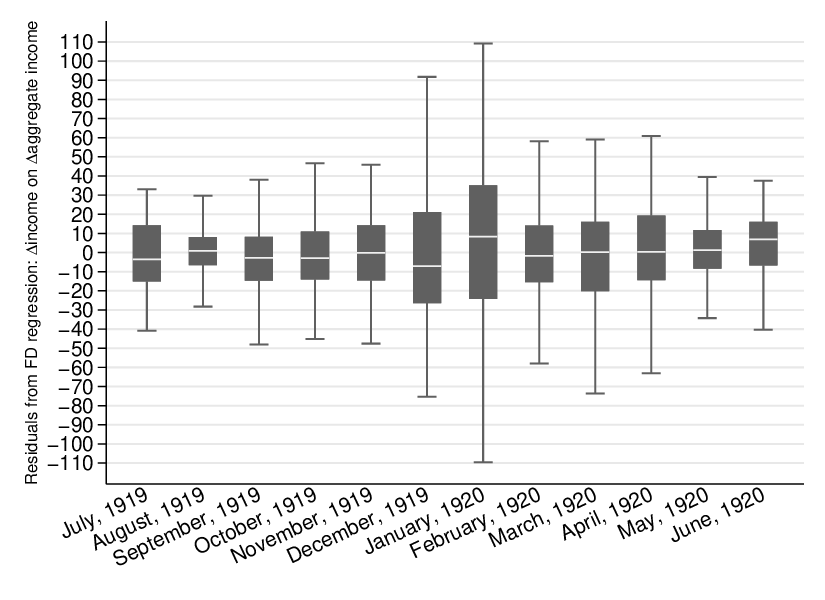}
\caption{Variations in idiosyncratic shocks (in yen)}
\label{fig:residuals}
\scriptsize{\begin{minipage}{350pt}
Notes: 
This figure shows the box-and-whisker plot of the residuals from the regression of the first-difference in income on the first-difference in aggregate income.
The 25th, 50th, and 75th percentiles are the bottom, middle (white line), and top of the box, respectively.
The caps at the ends of the whiskers show the lower and upper adjacent values, respectively.
Sources: Created by the author from the Municipal Bureau of Labor Research of Osaka (1919--1920).
\end{minipage}}
\end{figure}

Next, I examine the seasonality and variation of idiosyncratic shocks.
Figure~\ref{fig:mdeff} shows monthly income minus expenditure.
Seasonality is clearly observed in December 1919 and January 1920: both income and expenditure increase steeply in December, when bonuses have been paid (Tada 1991b, p.9).\footnote{This bonus was commonly paid to factory workers at that time (Bureau of Social Welfare 1923, p.~9). In Osaka, the bonus was roughly equivalent to one month's pay (Tada 1919a, pp.~40--49).}
Although the differences between income and expenditure were largely positive but fluctuating, this difference became statistically significant and negative in January because labor income and consumption expenditure typically increase in December to prepare for New Year events and traditions.
Workers also took New Year holidays, which could have reduced their earnings.
Figure~\ref{fig:residuals} illustrates the box-and-whisker plots of the residuals from the regression of the first-differences in household income on the first-differences in aggregate income, approximating the variations in idiosyncratic shocks during the sample period.\footnote{The specification for household $i$ in time $t$ can be written as $\Delta \textit{Income}_{i,t}= \alpha \Delta \overline{\textit{Income}}_{.,t}+ \Delta \eps_{i,t}$, where $\textit{Income}_{i,t}$ is monthly household income and $\overline{\textit{Income}}_{.,t}=n^{-1}_{n_{t}}\sum_{i=1}^{n_{t}}\textit{Income}_{i,t}$, in which $n_{t}$ indicates the number of households in time $t$.}
In the figure, the idiosyncratic shocks were clearer in those months in which the net income (i.e., income minus expenditure) exhibited large fluctuations (Figure~\ref{fig:mdeff}).
Further, the shocks were relatively large in March and April, that is, soon after the recession began.

\section{Consumption Smoothing} \label{sec:sec4}

\subsection{Estimation Strategy} \label{sec:sec41}

I begin my analysis by investigating whether risk was shared and which categories of consumption were robust to shocks.
Following Cochrane (1991) and Ravallion and Chaudhuri (1997), the empirical specification for household $i$ in time $t$ can be characterized as follows:
\begin{equation}\label{eqn:eq2}
\log c_{i,t} = \theta \log y_{i,t} + \mathbf{x'}_{i,g_{t}}\mathbf{\psi} + \mu_{i} + \phi_{t}+ u_{i,t},
\end{equation}
where $c_{i,t}$ is consumption, $y_{i,t}$ is disposable income, $\mu_{i}$ is the household fixed effect, $\phi_{t}$ is the month-year fixed effect, and $u_{i,t}$ is a random error term.\footnote{Online Appendix~\ref{sec:seca_theory1} describes the derivation of the specification. I use log disposable income as the income shock variable following the empirical setting in the literature (Cochrane 1991). Measurement error is unlikely to be problematic herein as the account books were rigorously checked by the investigators (Section~\ref{sec:sec32}). I confirm that my main results are driven from negative income shocks by analyzing an alternative specification using an indicator variable for the negative income shock. Online Appendix Table~\ref{tab:mace_rob} presents the results.}
Household-specific unobservables, such as permanent income and time-constant preferences, are captured by the household fixed effect. In addition, macroeconomic shocks and trends, including seasonality, are controlled for using the month-year fixed effect.
Evidence suggests that the households had not lost their heads but had maintained their size during the sample periods.
This means that unobservable preference shifts in household consumption do not disturb my results (see Online Appendix~\ref{sec:secb1} for details).\footnote{Lewis (1996) signaled that omitted variable bias may be an issue when the consumption of self-produced goods in village economies is ignored. However, since this study focuses on purchased consumption among urban households, this type of bias should be negligible.}
To be conservative, however, I consider the family size variables interacted with the quarter dummies to control for the potential preference shifts: $\mathbf{x}_{i,g_{t}}$ is a vector of the controls, where $g_{t} \in \{1, 2, 3, 4\}$ is the group membership variable indicating the quarters.
Note that the family size controls are interacted with the quarter dummies because these variables are surveyed once in the initial period (Panels A and C of Table~\ref{tab:sum} show the summary statistics).\footnote{
See, for instance, Naidu and Yuchtman (2013) for the validity of this strategy. The first quarter is set to be a reference quarter in all the regressions. I used the interactions with respect to quarter dummies rather than month-year dummies because optimization in the nonlinear fixed-effect model is unfeasible under such a complex model (Table~\ref{tab:mechanism}). I have confirmed that the results from the linear fixed-effect models, using the interaction terms with respect to year-month dummies, are virtually identical to the results from equation~\ref{eqn:eq2}.}
If idiosyncratic income shocks are perfectly insured, the coefficient on the change in the growth rate of individual income must be zero.
Therefore, the income elasticity captured as the estimate of $\theta$ ranges from zero (full insurance) to one (absence of insurance).

\subsection{Results} \label{sec:sec42}

To gain some insights on consumption smoothing, I begin by investigating the raw relationships between the changes in consumption and income.
Online Appendix Figures~\ref{fig:lnddinc} and \ref{fig:lndexp} show the distributions of the log-differences in monthly disposable income and expenditure, respectively.
Changes in the log of disposable income range from approximately $-2.5$ to $2.0$, whereas changes in the log of expenditure range from approximately $-1.5$ to $1.5$.
This finding suggests that income shocks are buffered to some degree.
An important fact here is that these idiosyncratic shocks are short-run (temporary) rather than long-run (persistent).
Indeed, most households that faced declining disposable income recovered the next month, suggesting that these shocks are caused by temporary sickness and/or layoffs as opposed to chronic diseases or long-term unemployment.
This feature of idiosyncratic shocks helps me examine the short-run responses of households.\footnote{In other words, to analyze households' responses to the long-term variations in incomes, much longer- and low-frequent-budget data are required (Attanasio and Davis 1996). Analyzing those historical budget datasets is another avenue for future work.}
Online Appendix Figure~\ref{fig:scat_exp} describes the relationship between the log-differences in monthly expenditure and disposable income, which shows a positive linear relationship.
To delve into this relationship, Figure~\ref{fig:scat_sub} decomposes total expenditure into the 10 subcategories in panel A in Table~\ref{tab:sum}, suggesting similar positive relationships between changes in income and most of those subcategories.
However, the relationships are rather unclear for housing and education expenditure, implying that some subcategories are less prone to being affected by income shocks.

Panel A of Table~\ref{tab:mace} presents the results for equation \ref{eqn:eq2}.
For the total consumption category, I find that the estimated coefficient is $0.39$ and is statistically significantly different from zero.
This implies that a one percent decrease in income results in a $0.39$ percent decrease in consumption.
Thus, this result suggests that the risks were not perfectly insured, but they might have been partially mitigated.
While comparable estimates for urban working-class households in developing economies are scarce, Townsend (1995) estimated an income elasticity of approximately $0.4$ in Bangkok in Thailand between 1975 and 1990.
Given that the average standard of living was in a similar range in both cases, this similarity is considered to be plausible.\footnote{Per-capita GDP (in 2011 dollars) in Japan in 1920 was $2,974$, whereas that in Thailand in 1975 was $3,123$ (Bolt and van Zanden 2020). While Thailand has experienced rapid economic growth since 1975, the comparison at the initial status should work because the increasing trend is captured by the year fixed effects in the model (Townsend 1995, p. 93).}
By contrast, Townsend (1994) found much smaller estimates (less than $0.14$) in Indian villages between 1975 and 1984.
Skoufias and Quisumbing (2005) also reported that the income elasticities of consumption were in a similar range in rural areas of Bangladesh, Ethiopia, and Mali in the 1990s (less than $0.15$).\footnote{Per-capita GDP (in 2011 dollars) in India was $1,430$ in 1975, whereas that in Bangladesh (in 1996), Ethiopia (in 1996), and Mali (in 1997) was also around $1,000$ (Bolt and van Zanden 2020).}
This implies that households in village economies are more likely to have access to informal insurance provided by networks, presumably through their familial relationships, than urban households.\footnote{Evidence from more developed societies in the 1990s also indicates that the income elasticity of total consumption in urban areas of Russia ($0.22$) was greater than that in rural areas of Mexico ($0.08$) (Skoufias and Quisumbing 2005). Per-capita GDP (in 2011 dollars) in Russia and Mexico was $12,400$ and $11,511$ in the initial survey years, respectively (Bolt and van Zanden 2020).}
However, directly comparing the income elasticity of urban economies with that of rural economies is practically difficult because village households can use self-consumption to mitigate idiosyncratic shocks, which tends to cause a downward bias in consumption.
Ravallion and Chaudhuri (1997) present much greater estimates (approximately $0.5$) by revising the bias in Townsend's (1994) estimates due to such measurement errors in consumption data.
This suggests that better access to savings and lending institutions, which is difficult for village households, can be important for mitigating idiosyncratic shocks.
While the urban-to-rural comparison of income elasticities is beyond the scope of this study, this debate indicates that research is required to understand risk-sharing behavior in developing economies.
In this light, the result provides the first benchmark estimate of income elasticity for any working-class household in the early phase of industrialization.

The estimates of the 10 subcategories in panel A of Table~\ref{tab:mace} are statistically significantly positive in most cases.
Overall, the elasticities are greater for ``luxury'' categories such as furniture ($0.66$), clothes ($0.68$), and entertainment expenses ($0.55$).
On the contrary, they are relatively small for the other subcategories: those for food, housing, education, medical expenses, and transportation are smaller than the elasticity for total consumption ($0.39$).
Specifically, as I show in a descriptive manner in Figure~\ref{fig:scat_sub}, the estimates of housing and education are particularly inelastic and even statistically insignificant at the critical level of 5\%.
The result for housing may not be surprising given that rent needs to be paid regardless of fluctuations.
Indeed, since most landlords of rental houses in cities leased land from landowners, they should have paid ground rent (Kato 1990, pp.~79--80).
This feature led to rental trouble for the working class around 1920, including evictions (Ono 1999, pp.~48--49).

By contrast, the result for education may need further clarification.
I first trim the sample to the $86$ households that have children aged 6--12.
As panel B-1 of Table~\ref{tab:mace} shows, the estimate for education is close to zero and statistically insignificant.
Next, I remove the potential impacts of graduation on education expenditure.
Because Japan's academic year starts in April and ends in March, some children might have graduated from school at the end of March 1920, which would disrupt education expenditure.
To address this issue, I run a regression using an alternative cut-off period between June 1919 and March 1920, as shown in panel B-2 of Table~\ref{tab:mace}.
In the same way, I exclude August 1919 to deal with potential fluctuations due to the summer holiday, as shown in Panel B-3 of Table~\ref{tab:mace}.
The results remain unchanged, suggesting that graduation and open seasonality do not affect the findings.

Mean monthly education expenses per child among RLR households with children aged 6--12 are $0.44$ yen, considerably lower than other expenses such as clothes, whose per-capita mean is $1.94$ yen in those households.
After the Order of Primary School (\textit{sh\=ogakk\=o rei}) was revised in 1900, primary school was regarded as compulsory education and tuition fees declined throughout the early 20th century (Hijikata 2002).
Indeed, average tuition spending was only $0.18$ yen per month in Osaka city in 1920, accounting for less than 0.2\% of the average monthly income of the households analyzed.\footnote{Most primary schools still collected tuition fees in Osaka city in the early 1920s even though they were no longer high (Osaka City Office 1922, pp.~3(8)--3(9); 3(22)--3(23)). The tuition fee was collected regularly on a specific day every month (Fuji 1925, p.~187). Other education expenses included smaller payments for school supplies such as stationery (Fuji 1925, pp.~90--91). This expenditure must also be included in the education expenses of RLR households.}
Consequently, withdrawals for economic reasons rarely occurred by the 1920s (Hijikata 2002, p.~32).\footnote{Even in the suburbs, the graduation rate from primary school was approximately 90\% (Okado 2000, p.~30). Hijikata (1994, p.~19) also provided evidence that the average dropout rate of primary schools in the 1920s was less than 10\%.}
This institutional advantage may explain my results.\footnote{As I explained earlier, the heads of RLR households were average factory workers (Section~\ref{sec:sec31}). By contrast, those children who could not attend school were mainly from low-income households whose heads were day laborers, rickshaw drivers, and handicraft makers at that time (Okado 2000, pp.~34--35).}
However, my result for education may be influenced by the data frequency: one can expect less insurance at the monthly level than annually because the consumption of some goods can be delayed in the short term (Nelson 1994).
In this light, education could be more of a long-term decision for parents.
Thus, one must be careful in concluding that RLR households kept children in education despite idiosyncratic shocks.

Nakagawa (1985) contended that urban factory worker households in the early 20th century kept spending on miscellaneous payments other than food to maintain their household (p.~379).
His argument is based on comparing expenditure on certain consumption categories calculated from several survey reports measured in different years.
Despite the methodological differences, my results for food and housing expenses are consistent with his argument.
Importantly, I also find that the income elasticities are positive in most consumption subcategories and that their magnitudes vary by category: payments for indispensable items such as food and housing are more likely to be inelastic than those for luxury items such as clothes and furniture.\footnote{Skoufias and Quisumbing (2005) found that food consumption is more likely to be insured from idiosyncratic shocks than nonfood consumption in Bangladesh, Ethiopia, Mali, Mexico, and Russia in the 1990s. While caution is required when comparing my results with those of other countries in the different developmental stages, the mechanism behind the adjustments must be similar. Another strand of the literature suggests that households could also alter their diet to adjust food expenditure (\"Oberg 2016).}

\section{Risk-Coping Strategies} \label{sec:sec5}

The foregoing results provide suggestive evidence that the factory worker households smoothed their consumption to some degree.
In this section, I investigate whether temporary sources of income such as savings and borrowing served to share risk among households and whether adjusting labor supply helped them cope with shocks.

\subsection{Savings and Borrowing} \label{sec:sec51}

\subsubsection*{\textit{Estimation Strategy}}

I derive the empirical specification to test risk-sharing strategies in the spirit of Fafchamps and Lund (2003).
The specification for household $i$ at time $t$ can be characterized as follows:
\begin{equation}\label{eqn:eq3}
r_{i,t} = \kappa + \delta \tilde{y}_{i,t} + \mathbf{x'}_{i,g_{t}} \vgamma + \nu_{i} + \zeta_{t} + e_{i,t},
\end{equation}
where $\tilde{y}_{i,t}$ is disposable income, $\vx_{i,g_{t}}$ is the family size controls included in equation~\ref{eqn:eq2}, $\nu_{i}$ indicates the household fixed effect, $\zeta_{t}$ indicates the month-year fixed effect, and $e_{i,t}$ is a random error term.\footnote{Online Appendix~\ref{sec:seca_theory2} describes the derivation of the specification. Disposable income in equation~\ref{eqn:eq3} excludes temporary income from borrowing, savings, and gifts as well as tax payments. I confirm that the results are largely unchanged if I use an indicator variable for negative shocks to disposable income instead of disposable income ($\tilde{y}_{i,t}$). Online Appendix~\ref{sec:secb2} summarizes these results.}
Since the household fixed effect absorbs variations in permanent income as well as the initial endowment of assets, disposable income ($\tilde{y}_{i,t}$) captures the idiosyncratic income shocks in the regression.

I use several income and expenditure categories for $r_{i,t}$.
First, temporary income from withdrawals and gifts is used to test the role of precautionary savings.
I also use monthly deposits to savings to confirm the saving behavior of the households: the signs of the estimates for the withdrawals and gifts and for deposits to savings should be symmetric if the households had precautionarily saved their earnings during good times.\footnote{In addition, the former estimate can be similar to or slightly greater than the latter estimate, as it might include a small gift amount.}
Second, I use temporary income from borrowing and liquidating loans to dually test the roles of lending institutions.
One can expect the estimates to be almost symmetric.
Finally, I use expenditure on clothes and furniture to test the potential role of pawnshops.
Although the borrowing category is indivisible, pawnshops were the most popular lending institutions and clothes were the dominant article for pawning (Section~\ref{sec:sec22}).
Hence, while the estimate for clothes could be sensitive to shocks, that for furniture, a placebo item, should be close to zero and statistically insignificant.

The analytical sample is trimmed to households that received income from either withdrawals and gifts or any borrowing during the sample period.\footnote{The results are materially similar if I use the full sample. However, I prefer to deal with the attenuation effects by removing the completely censored units. I have confirmed that the family size characteristics are similar in both the trimmed households and the rest. See Online Appendix~\ref{sec:secb3} for the finer details.}
Panels B-1 and B-2 of Table~\ref{tab:sum} list the summary statistics.

\subsubsection*{\textit{Results}}

Panels B-1 and B-2 of Table~\ref{tab:sum} indicate that households tended to rely more frequently on savings than borrowing ($1,023$ vs. $154$ obs.).
This suggests that households regard savings as the first line of defense and borrowing as the second line (Deaton 1991).
Given this, I begin my analysis by estimating income elasticity for the savings categories.

Panel A of Table~\ref{tab:mechanism} presents the results.
Columns (1) and (2) show the estimates for withdrawals and gifts.
Columns (3) and (4) show the estimates for deposits to savings.
The estimates in columns (1) and (3) suggest that when households faced negative income shocks, withdrawals and gifts increased, whereas deposits to savings decreased.\footnote{
To understand the saving behaviors of households in detail, I prefer to show the estimates for withdrawals and gifts, and deposits to savings separately. However, one can calculate the estimate for the net savings (say withdrawals minus deposits) as a combination of the estimates listed in columns (1) and (3): the elasticity of net savings is $-0.202-0.067=-0.269$. I provide the full results for net savings and net borrowing in Online Appendix Table~\ref{tab:mechanism_net}.}
To compare the estimate by addressing the attenuation effects induced by censoring, columns (2) and (4) employ the fixed-effects Tobit model proposed by Honor\'e (1992).
As expected, the magnitudes become larger in both columns.
The estimate for the withdrawals and gifts category is slightly higher than that for deposits to savings ($0.46$~vs.~$0.32$).
Although inconclusive, as the former income category is indivisible, this margin might suggest the potential contribution of gifts from informal insurance provided by networks (Rosenzweig 1988).

The estimate in column (2) suggests that a one standard deviation decrease in disposable income increases temporary income from withdrawals and gifts by $16$ yen, accounting for roughly 50\% of the average savings in savings banks of manufacturing workers in Osaka at that time (Section~\ref{sec:sec22}).
This finding implies that while precautionary savings served as a primary risk-coping strategy among households, they might not have provided sufficient compensation when households faced a rare but extreme loss of income (e.g., a more than two standard deviation decrease).
Despite this, the result for deposits to savings also evokes a serious attribute of urban factory worker households: they saved the surplus built up in good times to prepare for future risks.
This finding is consistent with James and Suto's (2011) historical view of frugal working-class households.


Panel B of Table~\ref{tab:mechanism} lists the estimates for the borrowing categories in the same column layout.
Columns (1) and (3) suggest that households borrowed money when they faced negative income shocks, whereas they liquidated loans in good times.
As expected, the magnitudes become much larger after addressing the censoring issues in columns (2) and (4).
To delve into the mechanism behind the roles of borrowing, I next examine the results for expenditure on luxury items listed in panel C of Table~\ref{tab:mechanism}: columns (1) and (2) for clothes and columns (3) and (4) for furniture.
While the estimates for clothes are weakly statistically significantly positive, the estimates from the placebo test using expenditure on furniture are close to zero and statistically insignificant.
This finding seems to be consistent with the historical fact that clothes were the most representative article for pawning at that time, whereas furniture was rarely used, suggesting a role for pawnshops in mitigating idiosyncratic income shocks (Section~\ref{sec:sec22}).
Fafchamps et al. (1998) showed that the ownership of mobile assets such as livestock can be used to mitigate vulnerability to idiosyncratic income shocks in village economies.
In this light, my result implies that urban factory worker households in prewar Japan owned clothes as real assets and bought some of those assets in good times, as suggested by Deaton (1992).

The estimate in column (2) of panel B of Table~\ref{tab:mechanism} implies that a one standard deviation decrease in disposable income increases temporary income from borrowing by $14$ yen.
Similarly, the estimate in column (4) indicates that a one standard deviation increase in disposable income increases liquidation by $12$ yen.
Given that the average amount borrowed by factory workers per event in pawnshops in Osaka city was $11$ yen, these magnitudes are quite reasonable (Section~\ref{sec:sec22}).
In other words, urban factory worker households took out loans within their solvencies.

\subsubsection*{\textit{Heterogeneous Responses}}

These findings suggest that both precautionary savings and borrowing mitigated idiosyncratic income shocks.
However, richer households with high precautionary savings may borrow less because they see consumption fluctuations with income as savings run out.\footnote{Deaton (1991) predicts the mechanism by which, among the households under a liquidity constraint, assets work to protect consumption against income shocks because the constraint increases the precautionary demand for savings. In Online Appendix Table~\ref{tab:mechanism_hetero_rob_const}, I provide suggestive evidence that precautionary savings might have mitigated the shocks among the households potentially facing borrowing constraints.}
To test such potential heterogeneous responses with respect to borrowing, I stratify the sample based on the median of households' mean savings expenses,\footnote{The results are unchanged if I use the expenditure on housing as an alternative cut-off variable that is exogenous to short-run earnings and savings. See Online Appendix~\ref{sec:secb5} for finer details of the discussion.} and run regressions for each subsample using equation~\ref{eqn:eq3}.\footnote{The specification using a product term between disposable income and a subsample dummy in equation~\ref{eqn:eq3} is undesirable because it postulates many considerably stronger assumptions on the remaining parameters, including the fixed effects.}
To deal with the attenuations from censoring revealed in Table~\ref{tab:mechanism}, I use the fixed-effects Tobit estimation in all the regressions.

Panel A of Table~\ref{tab:mechanism_hetero} presents the results.
Columns (1) and (3) show the estimates for withdrawals and gifts for the subsamples below and above the median, respectively.
The estimates are statistically significantly negative and have a similar range in both cases.
Columns (2) and (4) list the estimates of income from borrowing for the subsamples in the same layout.\footnote{Regressions using liquidation of loans are no longer practical under stratified subsamples because the optimization is computationally demanding, especially when the models are complex as in the case of small samples. Despite this, I confirm that the results for liquidation of loans from the simplified specification including disposable income and the household as well as the quarter-fixed effects are consistent with the results for income from borrowing: the estimates are statistically significantly positive in both subsamples, whereas the estimate for the subsample above the median is somewhat weak.}
The estimate for the subsample with less savings is statistically significantly negative, whereas the estimate for the subsample with more savings is negative but statistically insignificant.
The estimate in column (2) is greater than that in column (4) in an absolute sense.

Panel B of Table~\ref{tab:mechanism_hetero} assesses whether the results are consistent with the role of pawnshops, in the same manner as in panel C of Table~\ref{tab:mechanism}.
In this panel, I stratify the sample based on the median of households' mean monthly income from borrowing.
Columns (1) and (3) present the estimates for clothes for the subsamples below and above the median, respectively.
Income elasticity is statistically significantly positive among households that borrowed more, suggesting they might have stored the clothes in good times.
The estimate for households with less borrowing is higher, albeit statistically insignificant.
Although this result may partially reflect the income effects for clothes, the large standard error implies that a subset of households in this subsample did not pawn clothes.
The placebo results seem to support the potential mechanism via pawning, as shown in panel C of Table~\ref{tab:mechanism}, where the estimates for the furniture are closer to zero in both subsamples (columns (2) and (4)).

These results provide suggestive evidence that, while households regarded savings as a primary risk-coping strategy, borrowing was another option for relatively vulnerable households with less precautionary savings.

\subsection{Labor Supply Adjustments} \label{sec:sec52}

The increased dependence on the household head's earnings due to urbanization could increase demand for market purchases of insurance (di Matteo and Emery 2002).
However, additional labor supply could be an alternative means of coping with income shocks (Horrell and Oxley 2000; Moehling 2001).
In the foregoing regressions, I included family structure variables to control for potential preference shifts and related labor supply adjustments.
In this subsection, I explicitly test whether shocks to the income of the household head caused labor supply to be adjusted.

To create a suitable empirical setting for the test, I trim the sample to households with three or more family members, leaving $194$ households, including the head and both the wife and the child(ren).
I regress either the wife's or the child's income on the income of the household head, family size controls, and household and month-year fixed effects.\footnote{Income from the other family members (i.e., other than the head, wife, and child) is negligible because most of the observations in this category take zero values and thus any regressions are infeasible (Section~\ref{sec:sec32}).}
As shown in panel B-3 of Table~\ref{tab:sum}, approximately $20$\% ($347/1,627$) and $15$\% ($244/1,627$) of the observations have positive values for the wife's and child's incomes, respectively.
To deal with this censoring issue, I employ the fixed-effects Tobit estimation and run the wife's and child's income equations separately.\footnote{Neither the multinomial regression model nor the structural method proposed by Heckman (1979) are adopted because neither technique can profile out the household fixed effects. This is crucial in my analysis, which aims to estimate the income elasticity of the household head using the transitory part of that income after controlling for the household fixed effects. In addition, although the child's income is not divisible, such a scale effect is effectively controlled for by including the family size controls.}

Panel A of Table~\ref{tab:mechanism_ls} presents the results.
Column (1) shows the negative but insignificant estimate for the wife's income.
Column (2) shows the statistically significantly negative estimate for the child's income.\footnote{I confirm that this result is unchanged if I use the linear probability model for the indicator variables that take one for the observations with positive values of either the wife's or the child's income.}
The percentage of working children indicates that this result should come from children aged 13 years and older (i.e., those who have left primary school).\footnote{Although the ages of family members are unavailable, the RLR documented the number of family members in several age bins (panel C of Table~\ref{tab:sum}). Thus, I calculate the percentage of working children as the share of household-year-month cells with a positive income from the child to the total observations: the share is 25\% for households with children aged 13 and older, but only 1\% for households with children aged 12 or under.}
This finding is consistent with the discussion in Section~\ref{sec:sec42} about the low-income elasticity for education expenses.
In addition, it is in line with the findings from cross-sectional evidence of urban working-class households in prewar Japan exploiting the labor supply of children aged 15 years and older (Section~\ref{sec:sec22}).

However, my estimates from the subsamples stratified by the median of households' mean deposits to savings show a different story.
Panel B of Table~\ref{tab:mechanism_ls} presents the results.
Columns (1) and (2) show the results for the subsample below the median and columns (3) and (4) show those for the subsample above the median.
Among the former, the estimate for the wife's income is now statistically significantly negative (column (1)) and that for the child's income is higher (column (2)).
By contrast, the estimates are not statistically significant among households above the median (columns (3) and (4)).
Compared with the elasticities for savings and borrowing, which are approximately $0.5$ and $0.8$ (columns (1) and (2) in panel A of Table~\ref{tab:mechanism_hetero}), the elasticities of the labor supply of the wife and child (approximately $0.2$ and $0.6$, respectively) are slightly lower, but still within a similar range.
Therefore, adjusting labor supply could have been an important risk-coping strategy among households with less savings.
Finally, I confirm that these results are robust to using an alternative definition of the shock variable, namely, an indicator variable for the negative shock on the income of the household head (Online~Appendix~\ref{sec:secb2}).

\section{Conclusion}\label{sec:sec6}

Using a systematic empirical design with a unique household-based monthly-level panel dataset, this study is the first to investigate historical consumption smoothing behavior among factory worker households in an industrial city in Japan.
The income elasticity estimate of total consumption expenditure provides evidence that factory worker households in Osaka in the 1920s were as vulnerable to idiosyncratic shocks as urban households in the developing world at the same developmental stage, such as Bangkok during the 1970s and 1980s.
The estimated elasticities of the consumption subcategories also provide suggestive evidence that while households at that time could not fully cope with idiosyncratic shocks, they mitigated the fluctuations in indispensable consumption.
Indeed, the results of the mechanism analysis suggest that while households withdrew their savings as the first line of defense to deal with idiosyncratic income shocks, they saved the surplus in good times to prepare for future risks.
Among households with less savings, borrowing, particularly from pawnshops, worked as a second line of defense.
The additional labor supply of the wife and child in response to shocks to the income of the household head was also observed in relatively vulnerable households.

The risk-coping behavior among urban factory worker households in prewar Japan documented in this study indicates that they partially coped with idiosyncratic short-term income shocks using available sources of temporary income and adjusting labor supply.
To delve into the efficiency of historical insurance markets, it is also valuable to investigate whether the long-term variation in income was insured during historical economic growth.
As evidence from developed economies suggests that permanent shocks are far less insured than transitory shocks (Blundell et al.~2008), long-term shocks may rarely be insured in historical cases.
Despite this, the analyses based on the low-frequency budget data should provide new findings on the long-run risk-coping behaviors of the past.

Although the potential role of informal insurance, such as gifts from relatives and insurance sold in the market should be investigated in more detail, the results of this study reveal the risk-coping mechanisms and illustrate a much clearer picture of consumption smoothing behavior among urban working-class households than the traditional view.
This study provides the first estimates of income elasticities of consumption in the historical setting of working-class households in the first industrializing country in Asia.
Given the increasing availability of historical panel data, this study's estimates could serve as a benchmark for future studies on consumption smoothing behavior in other historical contexts.


\renewcommand{\refname}{Documents}

\newpage
\begin{figure}[]
\centering
\captionsetup{justification=centering}
  \subfloat[Food]{\label{fig:food}\includegraphics[width=0.2\textwidth]{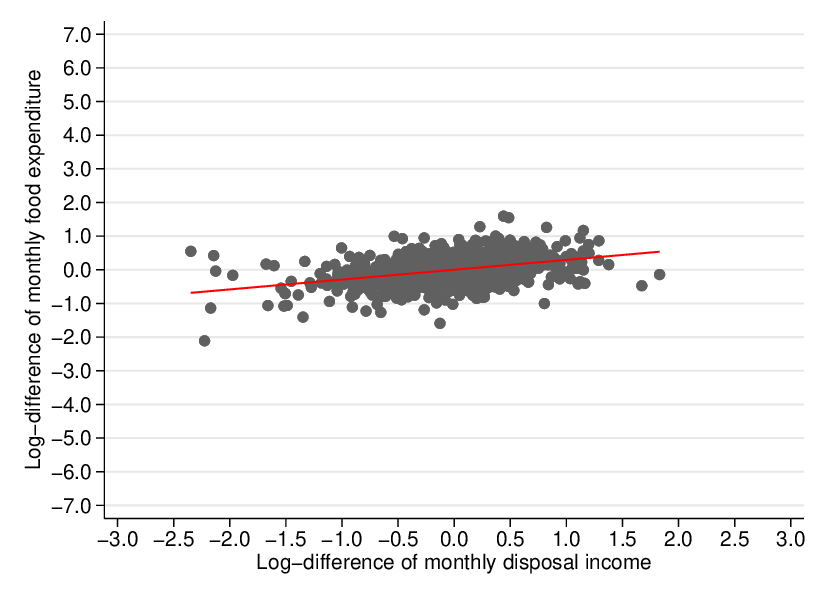}}
  \subfloat[Housing]{\label{fig:housing}\includegraphics[width=0.2\textwidth]{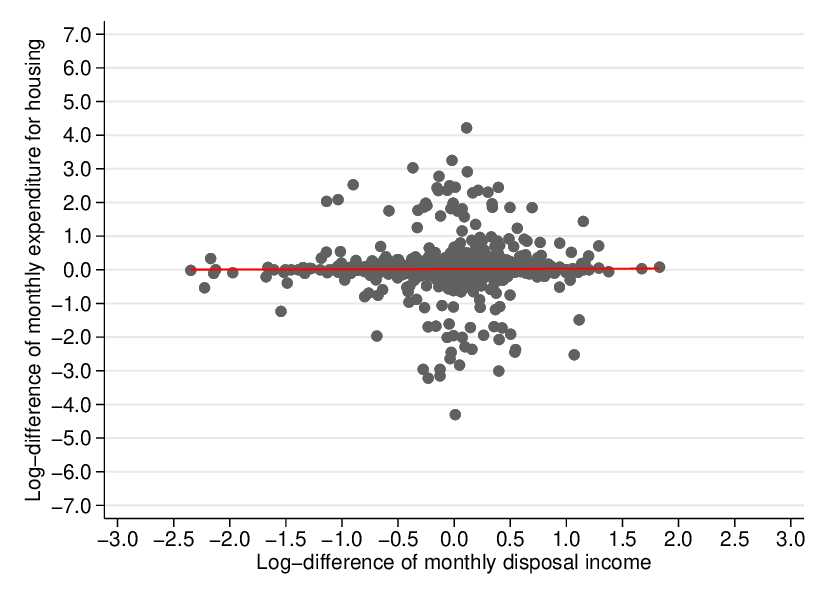}}
  \subfloat[Utilities]{\label{fig:utilities}\includegraphics[width=0.2\textwidth]{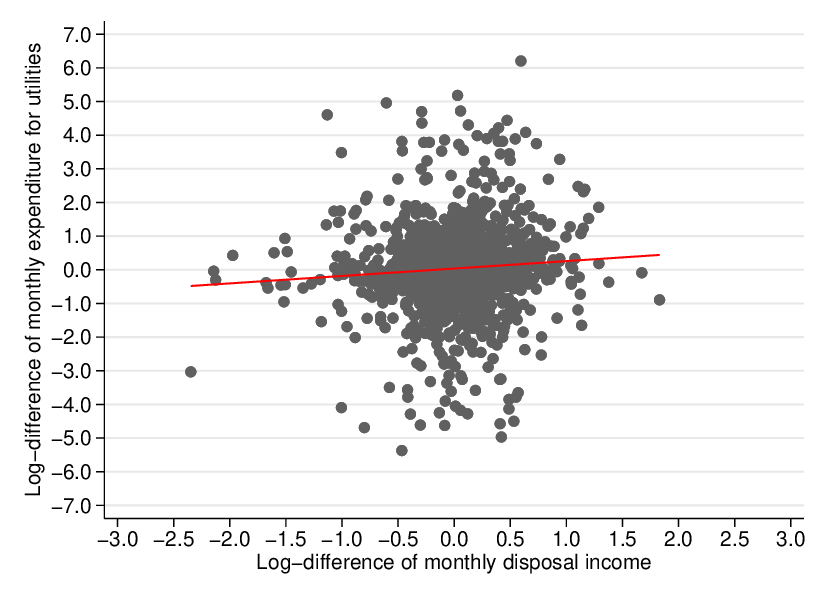}}
  \subfloat[Furniture]{\label{fig:furnishings}\includegraphics[width=0.2\textwidth]{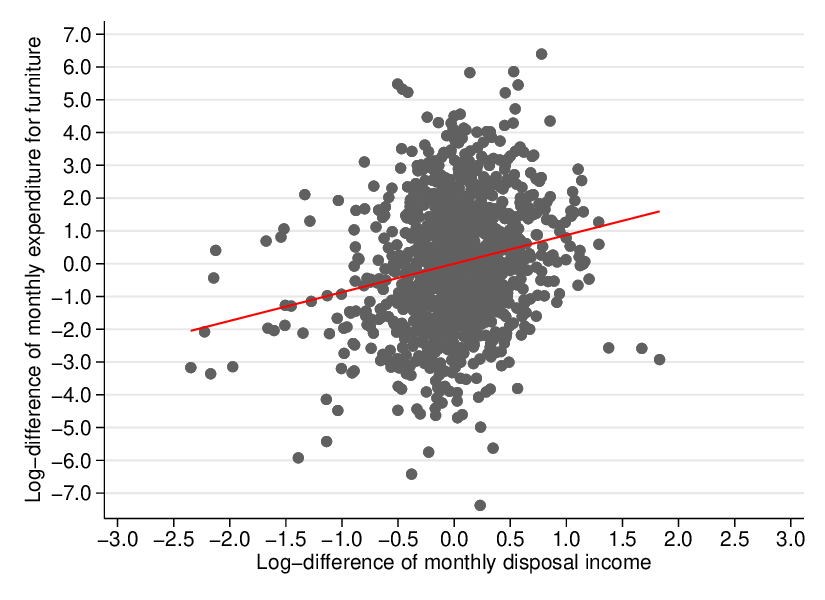}}
  \subfloat[Clothing]{\label{fig:clothing}\includegraphics[width=0.2\textwidth]{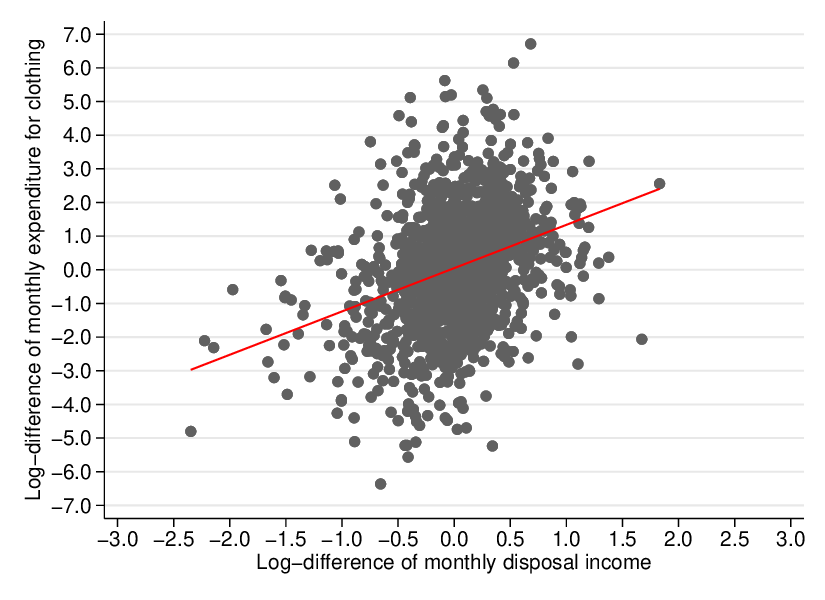}}\\
  \subfloat[Education]{\label{fig:education}\includegraphics[width=0.2\textwidth]{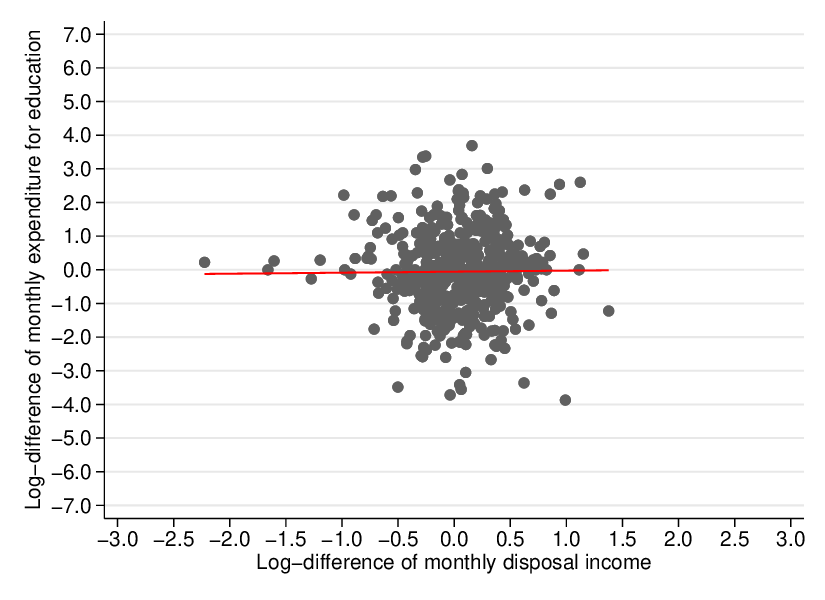}}
  \subfloat[Medical]{\label{fig:medical}\includegraphics[width=0.2\textwidth]{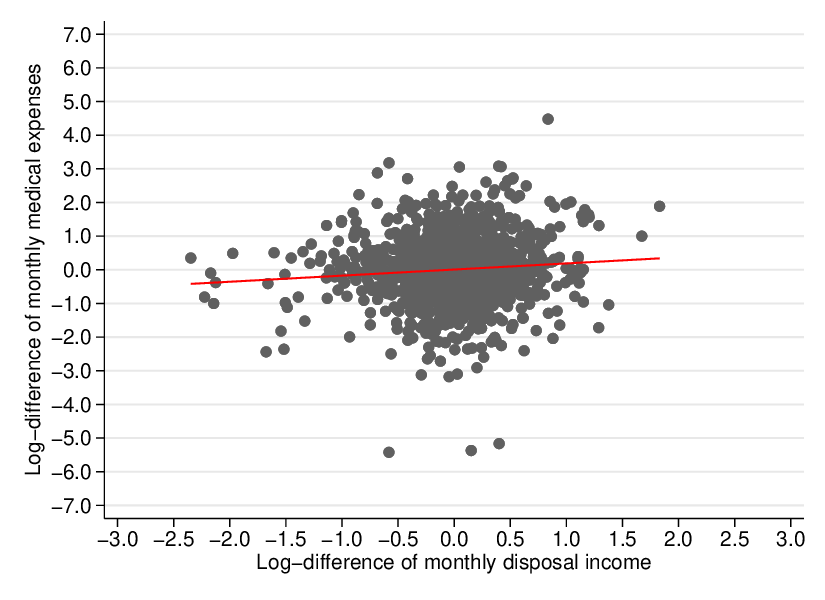}}
  \subfloat[Entertainment]{\label{fig:entertainment}\includegraphics[width=0.2\textwidth]{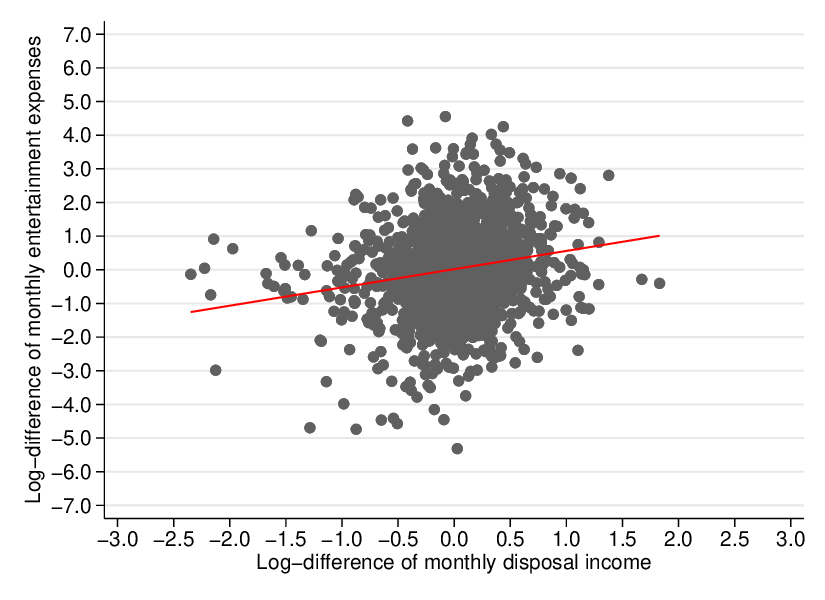}}
  \subfloat[Transportation]{\label{fig:transportation}\includegraphics[width=0.2\textwidth]{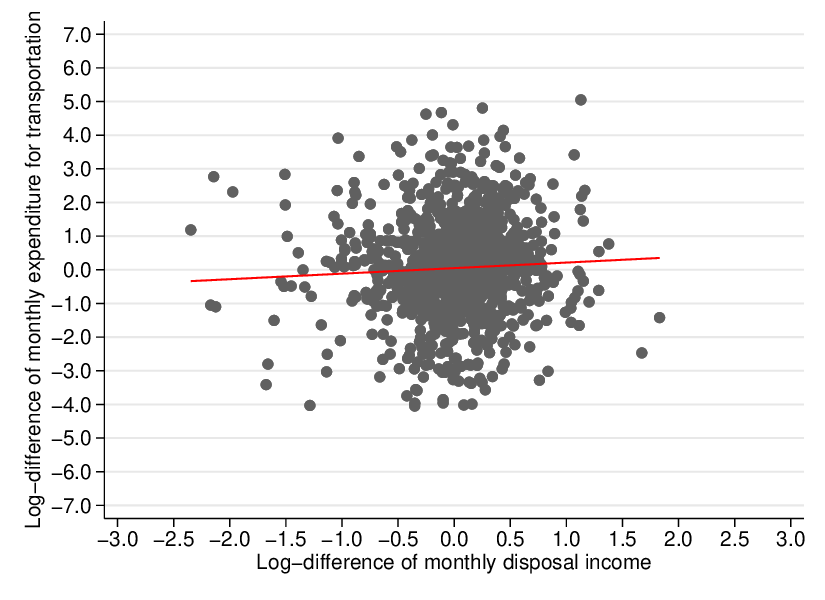}}
  \subfloat[Miscellaneous]{\label{fig:other}\includegraphics[width=0.2\textwidth]{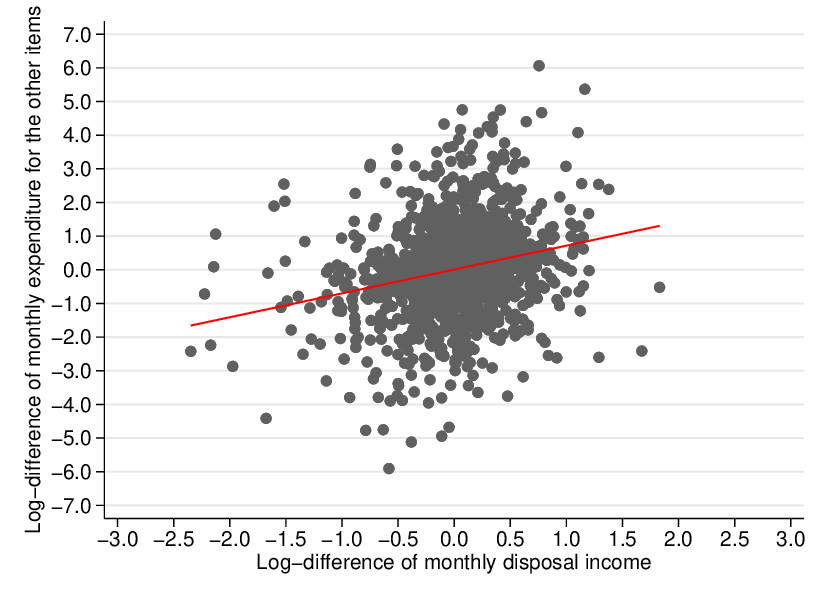}}
\caption{Relationship between the changes in disposable income and expenditure}
\label{fig:scat_sub}
\scriptsize{\begin{minipage}{450pt}
Notes:
The relationship between the changes in disposable income and expenditure for the 10 subcategories (panel A of Table~\ref{tab:sum}) is described in the figures.
For comparability, the minimum and maximum values of the y-axis are fixed at -7.0 and 7.0, respectively.
Sources: Created by the author from the Municipal Bureau of Labor Research of Osaka (1919--1920).
\end{minipage}}
\end{figure}

\clearpage
\begin{table}[!h]
\def\arraystretch{1.0}
\centering
\begin{center}
\caption{Sample Characteristics: Comparing the RLR Sample with the Census}
\label{tab:tab1}
\scriptsize
\scalebox{0.98}[1]{
\begin{tabular}{lrrrr}
\toprule
\multicolumn{5}{l}{\textbf{Panel A: Industrial structure}}\\
&&&&\\
\cmidrule(rrr){3-5}
\multicolumn{2}{l}{Name of survey}	&(1) 1920 Population census	&(2) 1920 Population census	&(3) The RLR\\
\multicolumn{2}{l}{Survey area}		&Osaka prefecture				&Osaka city						&Osaka city\\

\multicolumn{2}{l}{Survey subject}			&Complete survey&Complete survey&Sample from the city area\\
\multicolumn{2}{l}{Survey month and year}	&October 1920	&October 1920	&June 1919 to June 1920\\\hline

\multicolumn{2}{l}{Agriculture}						&14.1	&0.8	&0.0	\\ 
\multicolumn{2}{l}{Fisheries}						&0.5	&0.1	&0.0	\\
\multicolumn{2}{l}{Mining}							&0.3	&0.3	&0.0	\\
\multicolumn{2}{l}{Manufacturing}					&42.5	&45.6	&82.5	\\
\multicolumn{2}{l}{Commerce}						&25.8	&34.0	&2.0	\\
\multicolumn{2}{l}{Transport}						&8.7	&10.6	&4.2	\\
\multicolumn{2}{l}{Public service and professions}	&6.2	&6.7	&6.7	\\ 
\multicolumn{2}{l}{Housework}						&0.1	&0.1	&0.0	\\
\multicolumn{2}{l}{Other industry}					&1.8	&1.8	&4.4	\\\hline
&&&&\\
\multicolumn{5}{l}{\textbf{Panel B: Occupational structure and earnings}}\\
&&&&\\
\cmidrule(rrrr){2-5}
&\multicolumn{2}{c}{Manufacturing census in Osaka city}&\multicolumn{2}{c}{RLR sample}\\
\cmidrule(rr){2-3}\cmidrule(rr){4-5}
					&(1) \% share in	&(2) Monthly income of 	&(3) \% share in	&(4) Monthly income of\\
Sector				& adult male		&adult male (yen)		&heads			& heads (yen) [95\% CI]\\\hline
Textile				&12.5			&57.2					&8.9			&73.7 [64.3, 83.2]\\
Machine				&53.1			&70.7					&74.7			&69.6 [68.1, 71.0]\\
Chemical			&10.9			&65.6					&6.3			&76.1 [71.3, 80.8]\\
Food				&5.7			&62.2					&2.5			&62.3 [47.4, 77.2]\\
Miscellaneous		&17.8			&67.1					&7.6			&73.9 [64.7, 83.1]\\\bottomrule
\end{tabular}
}
{\scriptsize
\begin{minipage}{445pt}
Notes:\\
1. Panel A summarizes the industrial structures measured in the 1920 population census and RLR households.
Occupations are classified using the industrial classification of the first population census conducted in 1920.
The figures in column (3) are based on information on the 406 RLR household heads' occupations.
The five household heads whose occupations are classified as ``unknown'' are not included.\\
2. Panel B sorts the adult male factory workers in Osaka city in 1920 and heads of the 237 RLR households by the five sectors in the manufacturing industry.
Column (1) lists the percentage share of adult male factory workers in each sector.
Column (2) shows the mean monthly income of adult male factory workers.
The monthly income is calculated as the monthly earnings plus the monthly equivalent bonus.
See Online Appendix~\ref{sec:seca2} for the finer details.
Column (3) lists the percentage share of the household heads in each sector.
The sample excludes observations with no identifiable sector.
Column (4) shows the mean value of the household head's monthly income.\\
Sources:
Figures for the RLR households are calculated by the author from the Municipal Bureau of Labor Research of Osaka (1919--1920).
Industrial structures in panel A are from the Statistics Bureau of the Cabinet (1928, pp. 8--11) and (1929a, pp. 84--85, 108--109).
The occupational structure and earnings in panel B are from Osaka City Office (1921, pp.~8(44)--8(45); 1922, pp.~8(46)--8(47)).
\end{minipage}
}
\end{center}
\end{table}
\def\arraystretch{1.0}
\begin{table}[h]
\begin{center}
\caption{Summary Statistics: Sample of RLR Households}
\label{tab:sum}
\footnotesize
\scalebox{0.9}[1]{
\begin{tabular}{lrrrrrr}\toprule
\multicolumn{7}{l}{\textbf{Panel A: Variables for testing consumption smoothing}}\\
&\multicolumn{3}{c}{Raw variable}& \multicolumn{3}{c}{Log-transformed variable}\\
\cmidrule(rrr){2-4}\cmidrule(rrr){5-7}
&Mean&Std. dev.&Obs.&Mean&Std. dev.&Obs.\\\hline
Total consumption (yen)										&88.72	&38.54	&1880	&4.41	&0.38	&1880	\\
\hspace{10pt}Food											&41.83	&15.48	&1880	&3.67	&0.38	&1880	\\
\hspace{10pt}Housing										&7.75	&4.23	&1880	&1.91	&0.64	&1851	\\
\hspace{10pt}Utilities										&3.84	&2.86	&1880	&1.13	&0.94	&1733	\\
\hspace{10pt}Furniture										&2.22	&4.67	&1880	&0.04	&1.44	&1560	\\
\hspace{10pt}Clothes										&10.55	&15.22	&1880	&1.63	&1.34	&1840	\\
\hspace{10pt}Education										&0.49	&1.25	&1880	&-0.51	&1.15	&789	\\
\hspace{10pt}Medical expenses								&3.72	&4.37	&1880	&0.95	&0.86	&1866	\\
\hspace{10pt}Entertainment expenses						&5.42	&6.69	&1880	&1.19	&1.11	&1809	\\
\hspace{10pt}Transportation								&1.15	&1.85	&1880	&-0.32	&1.25	&1512	\\
\hspace{10pt}Miscellaneous									&5.47	&11.72	&1880	&1.03	&1.15	&1854	\\
Disposable income (yen)										&91.26	&41.51	&1880	&4.43	&0.41	&1880	\\\hline
&&&&&&\\
\multicolumn{7}{l}{\textbf{Panel B: Variables for testing risk-coping mechanisms}}\\
&\multicolumn{3}{c}{Analytical sample}& \multicolumn{3}{c}{Uncensored obs.}\\
\cmidrule(rrr){2-4}\cmidrule(rrr){5-7}
&Mean&Std. dev.&Obs.&Mean&Std. dev.&Obs.\\\hline
B-1: Variables for testing the role of savings (yen)		&&&&&&\\
\hspace{15pt}Withdrawals and gifts					&10.62	&22.92	&1,711	&17.76	&27.42	&1,023\\
\hspace{15pt}Deposits to savings					&3.83	&13.29	&1,711	&11.59	&21.10	&565	\\
\hspace{15pt}Disposable income					&80.18	&34.42	&1,711	&--&--&--\\
B-2: Variables for testing the role of borrowing (yen)	&&&&&&\\
\hspace{15pt}Borrowing							&5.09	&13.07	&599	&19.78	&19.38	&154	\\
\hspace{15pt}Liquidation of loans					&2.55	&7.32	&599	&11.16	&11.80	&137	\\
\hspace{15pt}Expenditure on clothes								&8.94	&11.17	&599	&9.10	&11.21	&588	\\
\hspace{15pt}Expenditure on furniture								&2.15	&5.02	&599	&2.55	&5.36	&506	\\
\hspace{15pt}Disposable income									&72.10	&27.77	&599	&--		&--		&--		\\
B-3: Variables for testing the labor supply adjustment					&&&&&&\\
\hspace{15pt}Wife's income											&2.60	&6.75	&1,627	&12.21	&9.83	&347	\\
\hspace{15pt}Child's income										&7.23	&21.14	&1,627	&48.22	&31.72	&244	\\
\hspace{15pt}Head's income										&70.63	&30.69	&1,627	&70.71	&30.60	&1,625	\\\hline
&&&&&&\\
\textbf{Panel C: Family size controls}&&&&&&\\
&Mean&Std. dev.&Min&Max&Obs.&\\
\cmidrule(lrrrrr){1-6}
Household size				&4.00	&1.61	&1	&9		&237&\\
Children aged 0--5 (\%)		&14.83	&16.02	&0	&60	&237&\\
Children aged 6--9 (\%)		&7.40	&11.26	&0	&40	&237&\\
Children aged 10--12 (\%)	&4.13	&9.09	&0	&40	&237&\\
Children aged 13--16 (\%)	&5.43	&10.38	&0	&50	&237&\\
Men aged 17+ (\%)			&33.52	&14.02	&0	&100	&237&\\
Women aged 17+ (\%)		&34.69	&15.72	&0	&100	&237&\\\bottomrule

\end{tabular}
}
{\scriptsize
\begin{minipage}{440pt}
Notes:\\
1.~Panel A: Summary statistics for all 237 RLR households. Disposable income is income excluding tax payments.\\
2.~Panel B: Summary statistics for the 202 households that received any income from withdrawals and gifts are listed in panel B-1.
The summary statistics for the 65 households that received any income from borrowing are listed in panel B-2.
Disposable income is income excluding tax payments, temporary income from borrowing, and withdrawals and gifts.
Summary statistics for the 194 households with three or more family members (i.e., households with children) are listed in panel B-3.\\
3.~Panel C: Summary statistics of the family size control variables for all 237 RLR households are listed.
The group of children aged 0--5 (\%) is used as the reference group in the regression analysis.\\
Sources: Calculated by the author from the Municipal Bureau of Labor Research of Osaka (1919--1920).
\end{minipage}
}
\end{center}
\end{table}
\def\arraystretch{1.0}
\begin{table}[]
\begin{center}
\caption{Results of Estimating Income Elasticities}
\label{tab:mace}

\footnotesize
\scalebox{1.0}[1]{
\begin{tabular}{lrlc}\toprule
&\multicolumn{2}{c}{Disposable income}&\\
\cmidrule(rl){2-3}
\textbf{Panel A: Main results}&Coef.&Std. error&Observations\\ \hline
\hspace{10pt}Total consumption 						&0.392	&[0.038]***	&1880  \\
&&&\\
\hspace{10pt}Food										&0.139	&[0.032]***	&1880  \\
&&&\\
\hspace{10pt}Housing									&0.081	&[0.044]*		&1851  \\
&&&\\
\hspace{10pt}Utilities									&0.293	&[0.105]***	&1733  \\
&&&\\
\hspace{10pt}Furniture									&0.656	&[0.163]***	&1560  \\
&&&\\
\hspace{10pt}Clothes									&0.682	&[0.113]***	&1840  \\
&&&\\
\hspace{10pt}Education									&0.091	&[0.129]		&789  \\
&&&\\
\hspace{10pt}Medical expenses							&0.381	&[0.076]***	&1866  \\
&&&\\
\hspace{10pt}Entertainment expenses					&0.554	&[0.095]***	&1809  \\
&&&\\
\hspace{10pt}Transportation							&0.305	&[0.128]**		&1512  \\
&&&\\
\hspace{10pt}Miscellaneous								&0.526	&[0.135]***	&1854  \\\hline
\\
&\multicolumn{2}{c}{Disposable income}&\\
\cmidrule(rl){2-3}
\textbf{Panel B: Additional results for education}&Coef.&Std. error&Observations\\ \hline
\multicolumn{4}{l}{B-1: Households with children aged 6--12}\\
\hspace{10pt}Education								&-0.014&[0.135]	&618   \\
\\
\multicolumn{4}{l}{B-2: Households with children aged 6--12 (June 1919 to March 1920)}\\
\hspace{10pt}Education								&0.033	&[0.147]	&466   \\
\\
\multicolumn{4}{l}{B-3: Households with children aged 6--12 (June 1919 to March 1920, excluding August)}\\
\hspace{10pt}Education								&0.079	&[0.155]	&450   \\
\bottomrule
\end{tabular}
}
{\scriptsize
\begin{minipage}{385pt}
***, **, and * denote statistical significance at the 1\%, 5\%, and 10\% levels, respectively.
Standard errors in brackets are clustered at the household level.\\
Notes: 
This table shows the results of equation~\ref{eqn:eq2}: the regressions of the 11 measures of log-transformed consumption on log-transformed disposable income as well as on the family size controls, household fixed effects, and month-year fixed effects.
The family size controls are interacted with the quarter dummies.
The estimated coefficients on log-transformed disposable income are listed in the second column (Coef.).
Panel A reports the estimates from the regressions using 237 households from June 1919 to June 1920.
Panel B-1 reports the estimates from the regression using the 86 households with children aged 6--12 from June 1919 to June 1920.
Panels B-2 and B-3 report the estimates from the regressions using the 82 households with children aged 6--12 from June 1919 to March 1920 and June 1919 to March 1920, excluding August 1919, respectively.
\end{minipage}
}
\end{center}
\end{table}
\def\arraystretch{1.0}
\begin{table}[]
\begin{center}
\caption{Results of Testing the Risk-Coping Mechanisms
\label{tab:mechanism}
}
\footnotesize
\scalebox{1.0}[1]{
\begin{tabular}{lrrrr}\toprule
\textbf{Panel A: Testing the role of savings}
&\multicolumn{4}{c}{Dependent variable}\\
\cmidrule(rrrr){2-5}
&\multicolumn{2}{c}{}&\multicolumn{2}{c}{Deposits to}\\
&\multicolumn{2}{c}{Withdrawals and gifts}&\multicolumn{2}{c}{savings}\\
\cmidrule(rr){2-3}\cmidrule(rr){4-5}
&\multicolumn{1}{c}{(1)}&\multicolumn{1}{c}{(2)}&\multicolumn{1}{c}{(3)}&\multicolumn{1}{c}{(4)}\\\hline
Disposable income							&-0.202***	&-0.457***	&0.067***	&0.324***	\\
											&[0.040]	&[0.097]	&[0.023]	&[0.087]	\\
Model										&Linear		&Nonlinear	&Linear		&Nonlinear	\\
Observations								&1,711		&1,711		&1,711		&1,711		\\\hline
&&&&\\
\textbf{Panel B: Testing the role of borrowing}
&\multicolumn{4}{c}{Dependent variable}\\
\cmidrule(rrrr){2-5}
&\multicolumn{2}{c}{}&\multicolumn{2}{c}{Liquidation of}\\
&\multicolumn{2}{c}{Borrowing}&\multicolumn{2}{c}{loans}\\
\cmidrule(rr){2-3}\cmidrule(rr){4-5}
&\multicolumn{1}{c}{(1)}&\multicolumn{1}{c}{(2)}&\multicolumn{1}{c}{(3)}&\multicolumn{1}{c}{(4)}\\\hline
Disposable income								&-0.122**	&-0.516***	&0.072*	&0.447***	\\
												&[0.047]	&[0.143]	&[0.039]	&[0.101]	\\
Model											&Linear		&Nonlinear	&Linear		&Nonlinear	\\
Observations									&599		&599		&599		&599		\\\hline
&&&&\\
\textbf{Panel C: Testing the role of pawnshops}
&\multicolumn{4}{c}{Dependent variable}\\
\cmidrule(rrrr){2-5}
&\multicolumn{2}{c}{Expenditure on}&\multicolumn{2}{c}{Expenditure on}\\
&\multicolumn{2}{c}{clothes}&\multicolumn{2}{c}{furniture}\\
\cmidrule(rr){2-3}\cmidrule(rr){4-5}
&\multicolumn{1}{c}{(1)}&\multicolumn{1}{c}{(2)}&\multicolumn{1}{c}{(3)}&\multicolumn{1}{c}{(4)}\\\hline
Disposable income								&0.076*	&0.110*	&0.014		&0.040		\\
												&[0.042]	&[0.058]	&[0.015]	&[0.037]	\\
Model											&Linear		&Nonlinear	&Linear		&Nonlinear	\\
Observations									&599		&599		&599		&599		\\\bottomrule
\end{tabular}
}
{\scriptsize
\begin{minipage}{415pt}
***, **, and * denote statistical significance at the 1\%, 5\%, and 10\% levels, respectively.
The results from the fixed-effects Tobit model, as proposed by Honor\'e (1992), are reported in columns (2) and (4) in each panel. 
A quadratic loss function is applied for the estimation to ensure computational tractability.
Robust standard errors are in brackets.
Standard errors are clustered at the household level in the linear models.\\
Notes:
Panel A presents the results for the savings category:
withdrawals and gifts (columns (1) and (2)) and deposits to savings (columns (3) and (4)).
Panel B presents the results for the borrowing category: borrowing (columns (1) and (2)) and liquidation of loans (columns (3) and (4)).
Panel C presents the results for expenditure on clothes (columns (1) and (2)) and furniture (columns (3) and (4)).
All the regressions in each panel include the disposable income, family size controls, household fixed effects, and month-year fixed effects.
The family size controls are interacted with the quarter dummies.
\end{minipage}
}
\end{center}
\end{table}
\def\arraystretch{1.0}
\begin{table}[]
\begin{center}
\caption{Results of Testing the Risk-Coping Mechanisms: Heterogeneous Responses
\label{tab:mechanism_hetero}
}
\footnotesize
\scalebox{0.94}[1]{
\begin{tabular}{lrrrr}\toprule
\multicolumn{5}{l}{\textbf{Panel A: Testing the role of savings and borrowing}}\\
&\multicolumn{4}{c}{Households' mean monthly deposits to savings}\\
\cmidrule(rrrr){2-5}
&\multicolumn{2}{c}{$\leq$ Median}&\multicolumn{2}{c}{$>$ Median}\\
\cmidrule(rr){2-3}\cmidrule(rr){4-5}
&\multicolumn{2}{c}{Income from}&\multicolumn{2}{c}{Income from}\\
\cmidrule(r){2-2}\cmidrule(r){3-3}\cmidrule(r){4-4}\cmidrule(r){5-5}
&\multicolumn{1}{c}{(1) withdrawals and gifts}&\multicolumn{1}{c}{(2) borrowing}&\multicolumn{1}{c}{(3) withdrawals and gifts}&\multicolumn{1}{c}{(4) borrowing}\\\hline
Disposable income							&-0.528***	&-0.841***	&-0.437***	&-0.262	\\
											&[0.076]	&[0.281]	&[0.144]	&[0.276]	\\
Observations								&855		&314		&856		&285		\\\hline
&&&&\\
\multicolumn{5}{l}{\textbf{Panel B: Testing the role of pawnshops}}		\\
&\multicolumn{4}{c}{Households' mean monthly income from borrowing}	\\
\cmidrule(rrrr){2-5}
&\multicolumn{2}{c}{$\leq$ Median}&\multicolumn{2}{c}{$>$ Median}	\\
\cmidrule(rr){2-3}\cmidrule(rr){4-5}
&\multicolumn{2}{c}{Expenditure on}&\multicolumn{2}{c}{Expenditure on}\\
\cmidrule(r){2-2}\cmidrule(r){3-3}\cmidrule(r){4-4}\cmidrule(r){5-5}
&\multicolumn{1}{c}{(1) clothes}&\multicolumn{1}{c}{(2) furniture}&\multicolumn{1}{c}{(3) clothes}&\multicolumn{1}{c}{(4) furniture}\\\hline
Disposable income								&0.143		&0.091		&0.077*	&0.061		\\
												&[0.087]	&[0.064]	&[0.040]	&[0.116]	\\
Observations									&333		&333		&266		&266		\\\bottomrule
\end{tabular}
}
{\scriptsize
\begin{minipage}{445pt}
***, **, and * denote statistical significance at the 1\%, 5\%, and 10\% levels, respectively.
The results from the fixed-effects Tobit model, as proposed by Honor\'e (1992), are reported. 
A quadratic loss function is applied for the estimation to ensure computational tractability.
Robust standard errors are in parentheses.\\
Notes: 
Panel A presents the results for the temporary income categories: columns (1) and (3) for withdrawals and gifts and columns (2) and (4) for borrowing.
The analytical sample used in panel A of Table~\ref{tab:mechanism_hetero} is stratified into two subsamples based on the median of households' mean monthly deposits to savings: columns (1) and (2) for households lower than or equal to the median (101 and 33 households for columns (1) and (2), respectively) and columns (3) and (4) for households higher than the median (101 and 32 households for columns (3) and (4), respectively).
Panel B presents the results for expenditure on clothes and furniture: columns (1) and (3) for clothes and columns (2) and (4) for furniture.
The analytical sample used in panel B of Table~\ref{tab:mechanism_hetero} is stratified into two subsamples by the median of households' mean monthly income from borrowing: columns (1) and (2) for the 31 households less than or equal to the median and columns (3) and (4) for the 34 households more than the median.
All the regressions in each panel include the disposable income, family size controls, household fixed effects, and month-year specific fixed effects.
The family size controls are interacted with the quarter dummies.
\end{minipage}
}
\end{center}
\end{table}
\def\arraystretch{1.0}
\begin{table}[]
\begin{center}
\captionsetup{justification=centering}
\caption{Results of Testing the Risk-Coping Mechanisms\\: Labor Supply Adjustments
\label{tab:mechanism_ls}
}
\footnotesize
\scalebox{1.0}[1]{
\begin{tabular}{lrrrr}\toprule
\multicolumn{5}{l}{\textbf{Panel A: Testing labor supply adjustments}}\\
&\multicolumn{4}{c}{Dependent variable}\\
\cmidrule(rrrr){2-5}
~~~~~~~~~~~~~~~~~~~~~~~~~~~~~~~~~~~
&\multicolumn{2}{c}{(1) Wife's income}&\multicolumn{2}{c}{(2) Child's income}\\\hline
Head's income		&\multicolumn{2}{c}{-0.059}	&\multicolumn{2}{c}{-0.410***}	\\
					&\multicolumn{2}{c}{[0.045]}	&\multicolumn{2}{c}{[0.111]}	\\
Observations		&\multicolumn{2}{c}{1,627}		&\multicolumn{2}{c}{1,627}		\\\hline
&&&&\\
\multicolumn{5}{l}{\textbf{Panel B: Testing the heterogeneous responses}}\\
&\multicolumn{4}{c}{Households' mean monthly deposits to savings}\\
\cmidrule(rrrr){2-5}
&\multicolumn{2}{c}{$\leq$ Median}&\multicolumn{2}{c}{$>$ Median}\\
\cmidrule(rr){2-3}\cmidrule(rr){4-5}
&\multicolumn{2}{c}{Income from}&\multicolumn{2}{c}{Income from}\\
\cmidrule(r){2-2}\cmidrule(r){3-3}\cmidrule(r){4-4}\cmidrule(r){5-5}
&\multicolumn{1}{c}{(1) wife}&\multicolumn{1}{c}{(2) child}&\multicolumn{1}{c}{(3) wife}&\multicolumn{1}{c}{(4) child}\\\hline
Head's income	&-0.186**	&-0.596***	&0.004		&-0.207	\\
					&[0.074]	&[0.155]	&[0.051]	&[0.224]	\\
Observations		&784		&784		&843		&843		\\\bottomrule
\end{tabular}
}
{\scriptsize
\begin{minipage}{340pt}
***, **, and * denote statistical significance at the 1\%, 5\%, and 10\% levels, respectively.
The results from the fixed-effects Tobit model, as proposed by Honor\'e (1992), are reported. 
A quadratic loss function is applied for the estimation to ensure computational tractability.
Robust standard errors are in parentheses.\\
Notes:
Panel A shows the results for the 194 households with three or more family members.
Panel B stratifies households based on the median of households' mean monthly deposits to savings.
Columns (1) and (2) show the results for the 97 households below the median and columns (3) and (4) present the results for the 97 households above the median.
All the regressions in each panel include the income of the household head, family size controls, household fixed effects, and month-year fixed effects.
The family size controls are interacted with the quarter dummies.
\end{minipage}
}
\end{center}
\end{table}
\clearpage
\thispagestyle{empty}

\begin{center}
\qquad

\qquad

\qquad

\qquad

\qquad

\qquad

{\LARGE \textbf{Appendices}}
\end{center}

\clearpage
\appendix
\def\thesection{Appendix~\Alph{section}}
\def\thesubsection{\Alph{section}.\arabic{subsection}}
\appendix
\def\thesection{Appendix~\Alph{section}}
\def\thesubsection{\Alph{section}.\arabic{subsection}}

\section{Theory Appendix}\label{sec:seca_theory}
\setcounter{page}{1}
\setcounter{figure}{0} \renewcommand{\thefigure}{A.\arabic{figure}}
\setcounter{table}{0} \renewcommand{\thetable}{A.\arabic{table}}

\subsection{Full-risk Sharing}\label{sec:seca_theory1}

Assume \textit{N} individuals named $i=1,..., N$ in the economy.
Individual \textit{i} receives an uncertain income $y_{i,t}(s_t)$, where $s_{t}\in S_{t}$ represents the state of the world at time \textit{t}, and derives instantaneous utility $u(c_{i,t}(s_{t}), h_{i,t}(s_{t}))$ from consumption $c_{i,t}(s_{t})$. 
The weighted sum of expected lifetime utility of $N$ individuals is expressed as:
\begin{equation}\label{eqn:of}
\sum_{i=1}^N\omega_{i}\sum_{t=0}^\infty \beta^t\sum_{s_{t}\in S_{t}} \pi(s_{t})u(c_{i,t}(s_{t}), h_{i,t}(s_{t})), 
\end{equation}
where $\omega_{i}$ is the social planner's weight, which is the reciprocal of the marginal utility of each agent, and satisfies $0<\omega_{i}<1$ (Negishi 1960); $0<\beta^t<1$ is the discount factor; $\pi(s_{t})\in [0, 1]$ is the probability that state $s_{t}$ takes place at time \textit{t}; and $h_{i,t}(s_{t})$ is a preference shock. 
The social planner maximizes the objective function (\ref{eqn:of}) by choosing an allocation of consumption across individuals, subject to the aggregate resource constraint of the form: 
\begin{equation}\label{eqn:rc}
\sum_{i=1}^N c_{i,t}(s_{t})=\sum_{i=1}^N y_{i,t}(s_{t}).
\end{equation}
Postulating a constant absolute risk aversion preference, $u(c_{i,t}(s_{t}), h_{i,t}(s_{t}))=-\frac{1}{\sigma}\exp(-\sigma(c_{i,t}(s_{t})-h_{i,t}(s_{t})))$, where $\sigma>0$ is the coefficient of constant absolute risk aversion, 
I can obtain the first-order condition for individual \textit{i}:
\begin{equation}\label{eqn:foc}
\omega_{i}\beta^t\pi(s_{t})\exp(-\sigma(c_{i,t}(s_{t})-h_{i,t}(s_{t})))=\lambda (s_{t}), 
\end{equation}
where $\lambda (s_{t})$ is the Lagrange multiplier for the resource constraint (\ref{eqn:rc}) at time \textit{t}. 
Taking the log of equation (\ref{eqn:foc}) and aggregating over agents, I obtain individual \textit{i}'s consumption as follows: 
\begin{equation}\label{eqn:cst1}
c_{i,t}(s_{t})=\frac{1}{N} \sum_{i=1}^N c_{i,t}(s_{t})+\frac{1}{\sigma}\Bigl(\log \omega_{i}-\frac{1}{N} \sum_{i=1}^N \log \omega_{i}\Bigr)+h_{i,t}(s_{t})-\frac{1}{N} \sum_{i=1}^N h_{i,t}(s_{t}).
\end{equation}
For simplicity, I use the conventional notation for a random variable $c_{i,t}\equiv c_{i,t}(s_{t})$, $h_{i,t}\equiv h_{i,t}(s_{t})$, and $\lambda_{t} \equiv \lambda (s_{t})$. 
Finally, equation (\ref{eqn:cst1}) with this notation becomes
\begin{equation}\label{eqn:cst2}
c_{i,t}=c_{t}^{a}+\frac{1}{\sigma}(\log \omega_{i}-\omega^{a})+(h_{i,t}-h_{t}^{a}), 
\end{equation}
where
\begin{equation}\label{eqn:agg}
c_{t}^{a}=\frac{1}{N} \sum_{i=1}^N c_{i,t},~~~~~\omega^a=\frac{1}{N} \sum_{i=1}^N \log \omega_{i},~~~~~h_{t}^{a}=\frac{1}{N} \sum_{i=1}^N h_{i,t}.
\end{equation}
The first-difference in equation (\ref{eqn:cst2}) eliminates the individual fixed effects to yield
\begin{equation}\label{eqn:fd}
c_{i,t}-c_{i,t-1}=c_{t}^{a}-c_{t-1}^{a}+h_{i,t}-h_{i,t-1}-(h_{t}^{a}-h_{t-1}^{a}). 
\end{equation}
Using the change in individual income $y_{i,t}-y_{i,t-1}$ as a proxy for idiosyncratic shocks, the empirical specification can then be characterized as
\begin{equation}\label{eqn:reg1}
c_{i,t}-c_{i,t-1}=\alpha_1(c_{t}^{a}-c_{t-1}^{a})+\alpha_2(y_{i,t}-y_{i,t-1})+\epsilon_{i,t}, 
\end{equation}
where $\epsilon_{i,t}$ is a disturbance term that includes both the time-varying preference shock, which affects individual-level consumption, and measurement errors in the data.
Equation (\ref{eqn:reg1}) in the case of a constant relative risk aversion preference can also be expressed as
\begin{equation}\label{eqn:reg2}
\log c_{i,t}- \log c_{i,t-1}=\alpha_1(\log c_{t}^{a}-\log c_{t-1}^{a})+\alpha_2(\log y_{i,t}-\log y_{i,t-1})+\epsilon_{i,t},
\end{equation}
where $\log (c_{i,t} / c_{i,t-1})$, $\log (c_{t}^{a} / c_{t-1}^{a})$, and $\log (y_{i,t} / y_{i,t-1})$ are the growth rates of individual consumption, aggregate consumption, and individual income, respectively.

This growth specification (\ref{eqn:reg2}) assumes that the aggregate measure of consumption captures macroeconomic shocks.
However, aggregate consumption in my dataset is for households in the manufacturing industry, making the interpretation of $\alpha_1$ problematic.
To address this issue, I use a two-way fixed-effects model instead of the first-difference model.
Following Cochrane (1991) and Ravallion and Chaudhuri (1997), the empirical specification can be simplified as follows:
\begin{equation}\label{eqn:reg3}
\log c_{i,t} = \theta \log y_{i,t} + \mathbf{x'}_{i,g_{t}}\mathbf{\psi} + \mu_{i} + \phi_{t}+ u_{i,t},
\end{equation}
where $c_{i,t}$ is consumption, $y_{i,t}$ is disposable income, $\mathbf{x}_{i,g_{t}}$ is a vector of the controls (Section~\ref{sec:sec41}), $\mu_{i}$ is the household fixed effect, $\phi_{t}$ is the month-year fixed effect, and $u_{i,t}$ is a random error term.
If risk is fully shared among individuals, the coefficient on the change in the growth rate of individual income becomes zero.
Hence, one can surmise that the estimate of $\theta$ range from zero (for full-risk sharing, where idiosyncratic shocks are perfectly insured) to one (for the absence of insurance).

\subsection{Risk-coping Strategies}\label{sec:seca_theory2}

To test the risk-coping mechanism, I consider that household consumption can be defined in the spirit of Fafchamps and Lund (2003):
\begin{equation}\label{eqn:consumption2}
c_{i,t}=y_{i}^{P} + y_{i,t}^{T} + z_{i,t} + b_{i,t},
\end{equation}
where $z_{i,t}$ and $b_{i,t}$ indicate net income from withdrawals and gifts and borrowing, respectively.
$y_{i}^{P}$ and $y_{i,t}^{T}$ are permanent and transitory income, respectively.
Equation (\ref{eqn:cst2}) can then be rewritten by substituting equation (\ref{eqn:consumption2}):
\begin{equation}\label{eqn:consumption3}
z_{i,t} + b_{i,t} = - (y_{i}^{P} + y_{i,t}^{T}) + c_{t}^{a}+\frac{1}{\sigma}(\log \omega_{i}-\omega^{a})+(h_{i,t}-h_{t}^{a}).
\end{equation}
The time-constant components ($y_{i}^{P}$, $\omega_{i}$) and individual-constant components ($c_{t}^{a}$, $h_{t}^{a}$) can be replaced by the individual fixed effects ($\nu_{i}$) and time fixed effects ($\lambda_{t}$), respectively.
Transitory income and preference shocks ($y_{i,t}^{T}$, $h_{i,t}$) can be replaced by disposable income ($\tilde{y}_{i,t}$) and the observable family characteristics ($\vx_{i,g_{t}}$), respectively.
Under these assumptions, the empirical specification is as follows:
\begin{equation}\label{eqn:consumption4}
z_{i,t} + b_{i,t} = \kappa + \delta \tilde{y}_{i,t} + \mathbf{x'}_{i,g_{t}} \vgamma + \nu_{i} + \zeta_{t} + e_{i,t},
\end{equation}
where $e_{i,t}$ is a random error term.
I regress withdrawals and gifts ($z_{i,t}$) and borrowing ($b_{i,t}$) separately on the shock variable (Section~\ref{sec:sec51}).
If idiosyncratic shocks are compensated by the risk-coping strategies, the estimated coefficient on the shock variable should be negative and statistically significant.

\section{Data Appendix}\label{sec:seca}
\setcounter{page}{1}
\setcounter{footnote}{0}
\setcounter{figure}{0} \renewcommand{\thefigure}{B.\arabic{figure}}
\setcounter{table}{0} \renewcommand{\thetable}{B.\arabic{table}}

\subsection{Monthly Income and Expenditure}\label{sec:seca1}

\begin{figure}[h]
\centering
\captionsetup{justification=centering}
  \subfloat[Disposable income]{\label{fig:dinc}\includegraphics[width=0.33\textwidth]{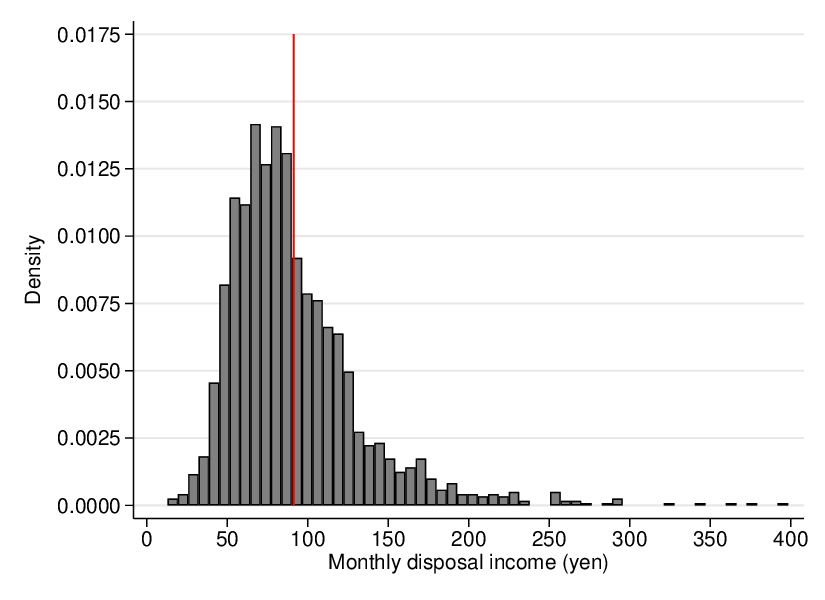}}
  \subfloat[Expenditure]{\label{fig:exp}\includegraphics[width=0.33\textwidth]{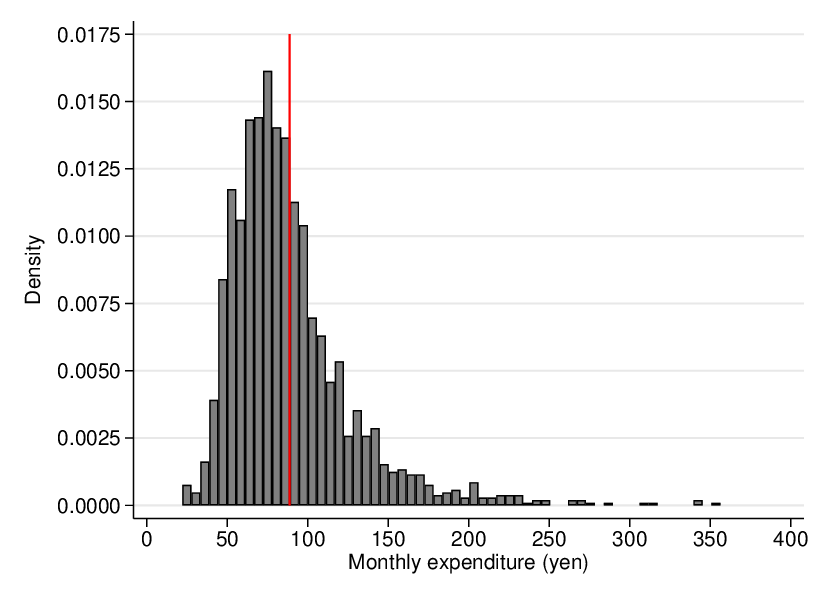}}
  \subfloat[Correlation]{\label{fig:scat_expinc}\includegraphics[width=0.33\textwidth]{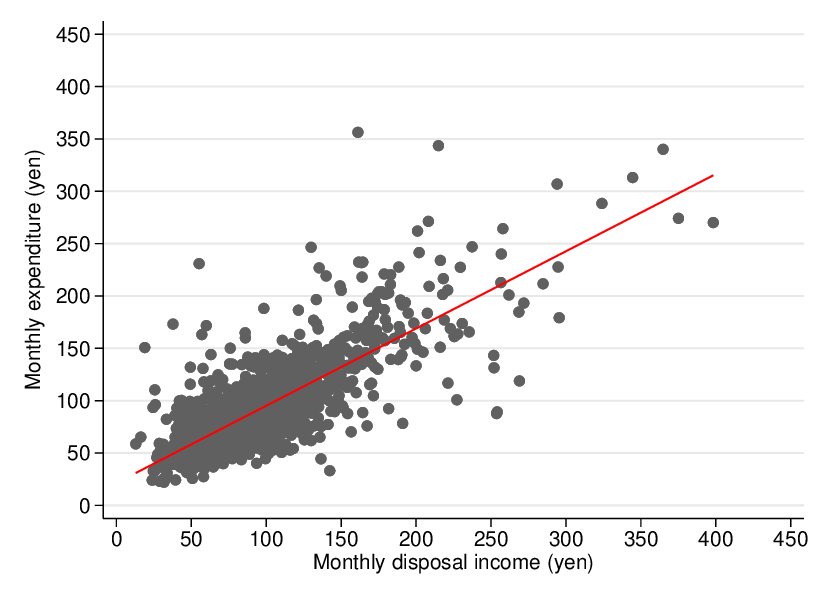}}
\caption{Disposable income and expenditure (yen)}
\label{fig:dincexp}
\scriptsize{\begin{minipage}{450pt}
Notes:
Monthly disposable income and expenditure are illustrated in the figures.
Disposable income is income excluding tax payments.
The solid red lines in Figure~\ref{fig:dinc} and \ref{fig:exp} indicate mean monthly disposable income and expenditure, respectively.
Figure~\ref{fig:scat_expinc} illustrates the correlation between monthly disposable income and monthly expenditure.
Sources: Created by the author from the Municipal Bureau of Labor Research of Osaka (1919--1920).
\end{minipage}}
\end{figure}

\begin{figure}[]
\centering
\captionsetup{justification=centering}
  \subfloat[Disposable income]{\label{fig:lnddinc}\includegraphics[width=0.34\textwidth]{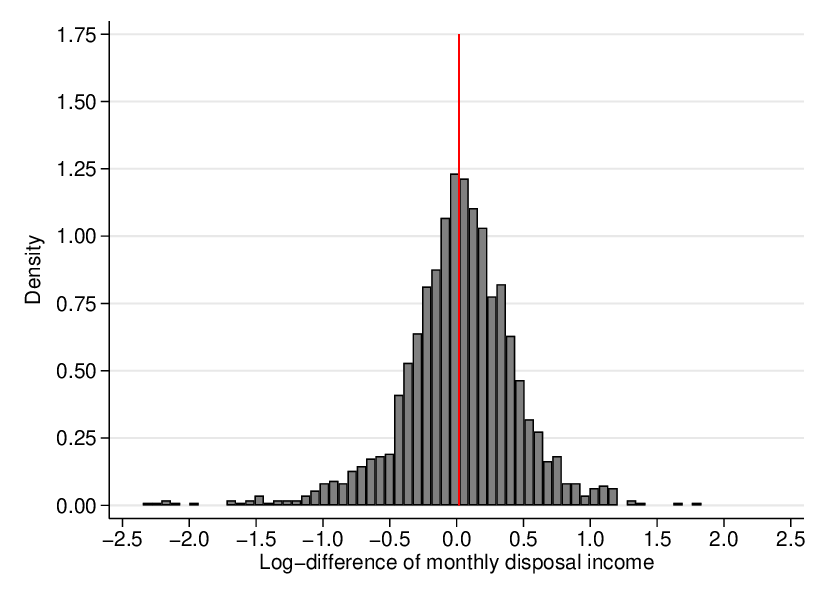}}
  \subfloat[Expenditure]{\label{fig:lndexp}\includegraphics[width=0.34\textwidth]{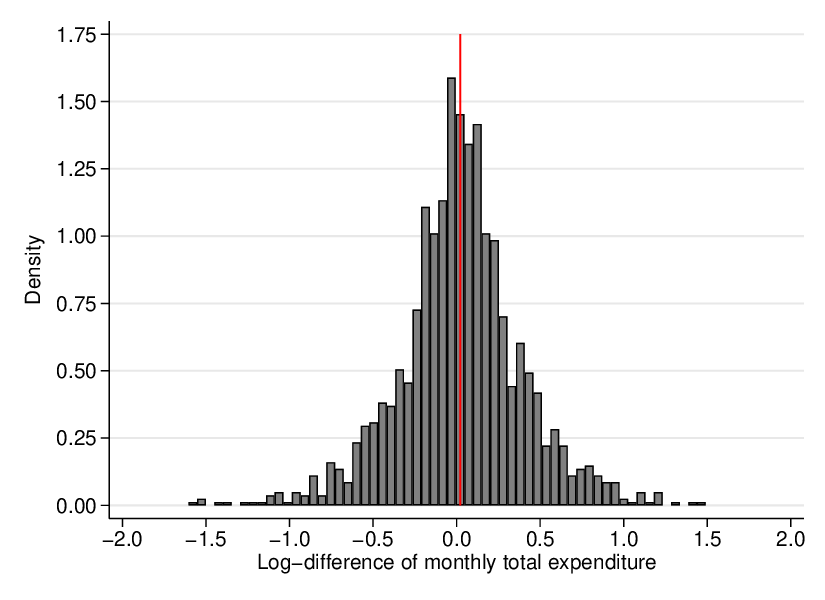}}
  \subfloat[Correlation]{\label{fig:scat_exp}\includegraphics[width=0.34\textwidth]{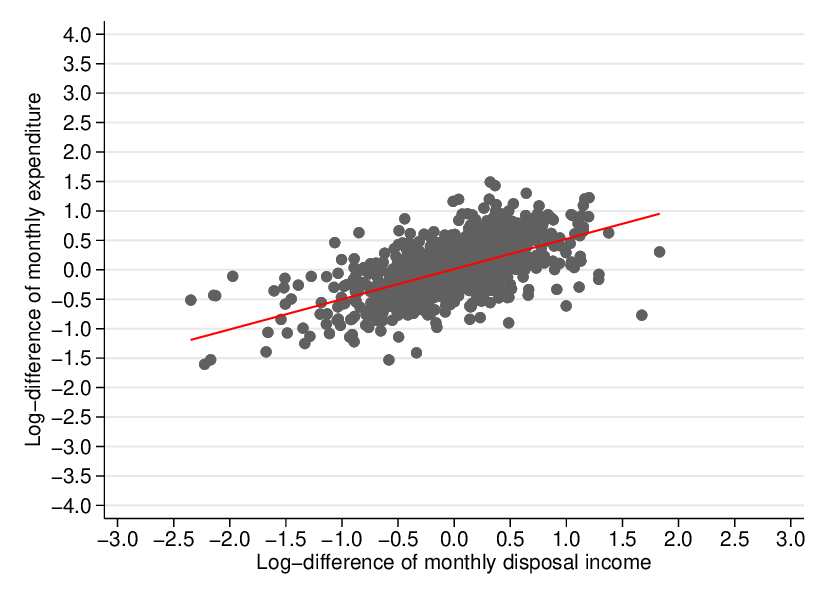}}
\caption{Log-difference in disposable income and expenditure}
\label{fig:lddincexp}
\scriptsize{\begin{minipage}{450pt}
Notes:
Distributions of the log-differences in disposable income and expenditure are shown in Figures~\ref{fig:lnddinc} and \ref{fig:lndexp}, respectively.
Disposable income is income excluding tax payments.
The solid red lines in Figures~\ref{fig:lnddinc} and \ref{fig:lndexp} indicate the mean of the log-differences in monthly disposable income and expenditure, respectively.
Figure~\ref{fig:scat_exp} describes the correlation between the log-differences in monthly disposable income and expenditure.
Sources: Created by the author from the Municipal Bureau of Labor Research of Osaka (1919--1920).
\end{minipage}}
\end{figure}

\begin{figure}[h]
\centering
\captionsetup{justification=centering}
  \subfloat[Food]{\label{fig:histfood}\includegraphics[width=0.2\textwidth]{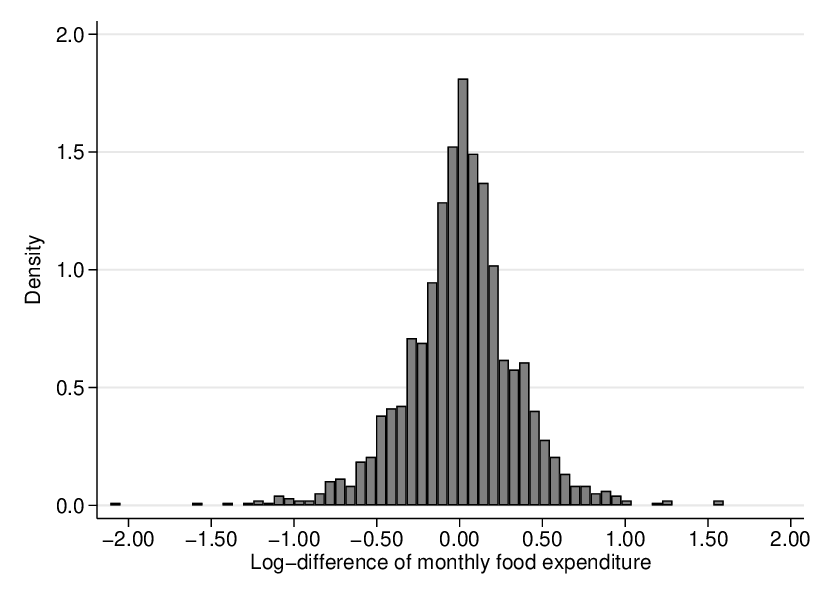}}
  \subfloat[Housing]{\label{fig:histhousing}\includegraphics[width=0.2\textwidth]{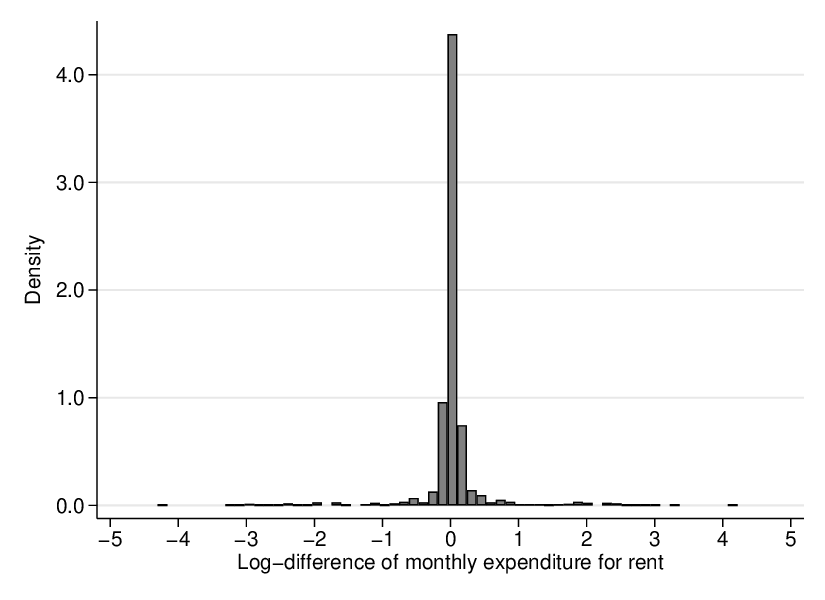}}
  \subfloat[Utilities]{\label{fig:histutilities}\includegraphics[width=0.2\textwidth]{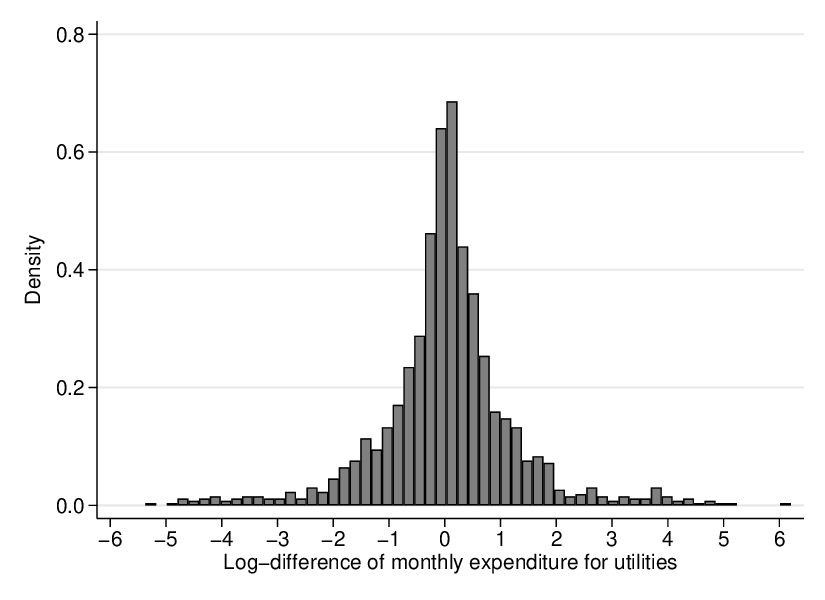}}
  \subfloat[Furniture]{\label{fig:histfurnishings}\includegraphics[width=0.2\textwidth]{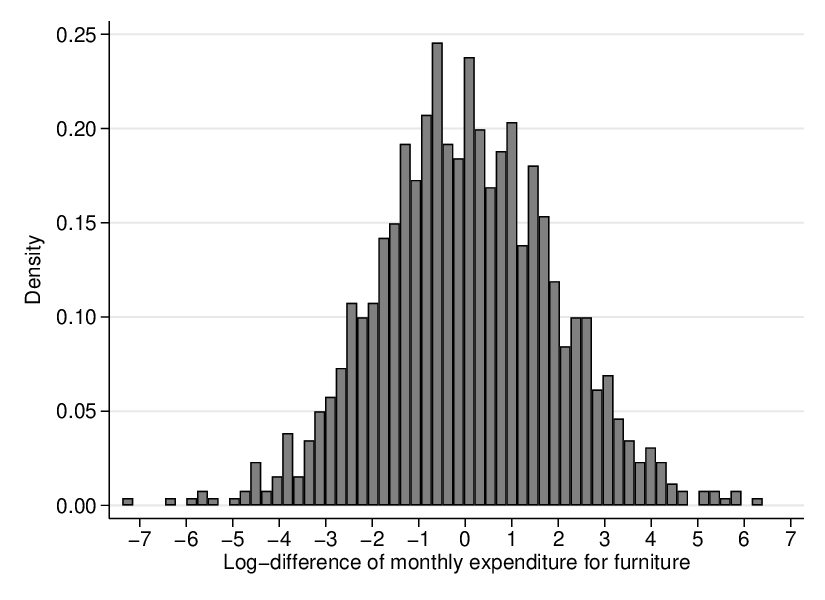}}
  \subfloat[Clothing]{\label{fig:histclothing}\includegraphics[width=0.2\textwidth]{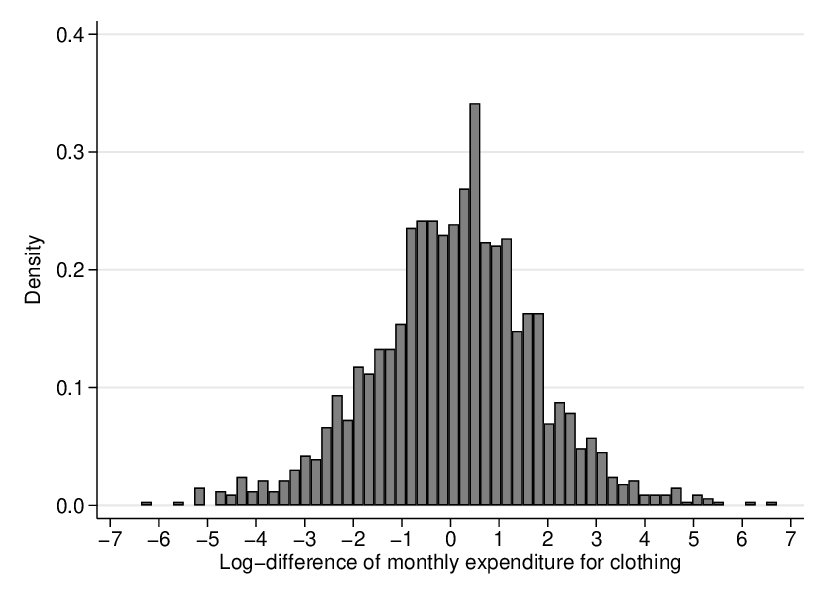}}\\
  \subfloat[Education]{\label{fig:histeducation}\includegraphics[width=0.2\textwidth]{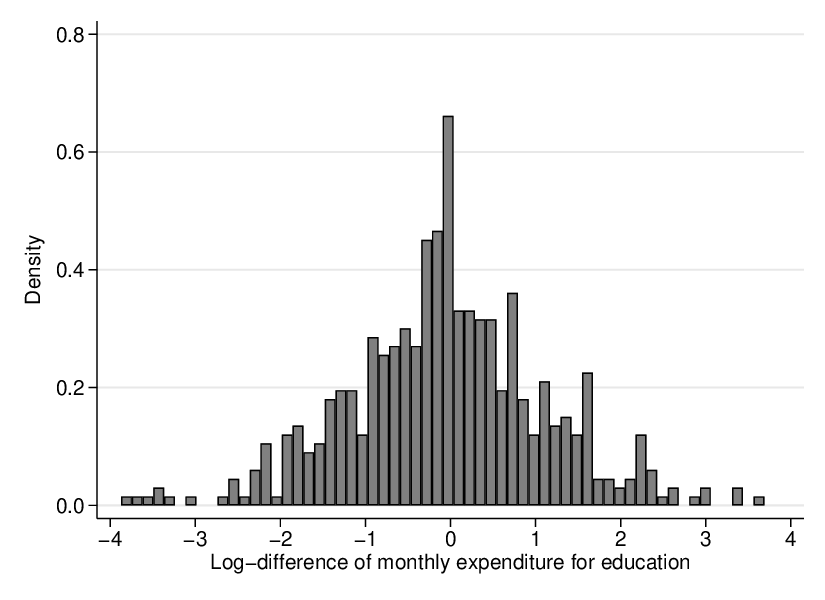}}
  \subfloat[Medical]{\label{fig:histmedical}\includegraphics[width=0.2\textwidth]{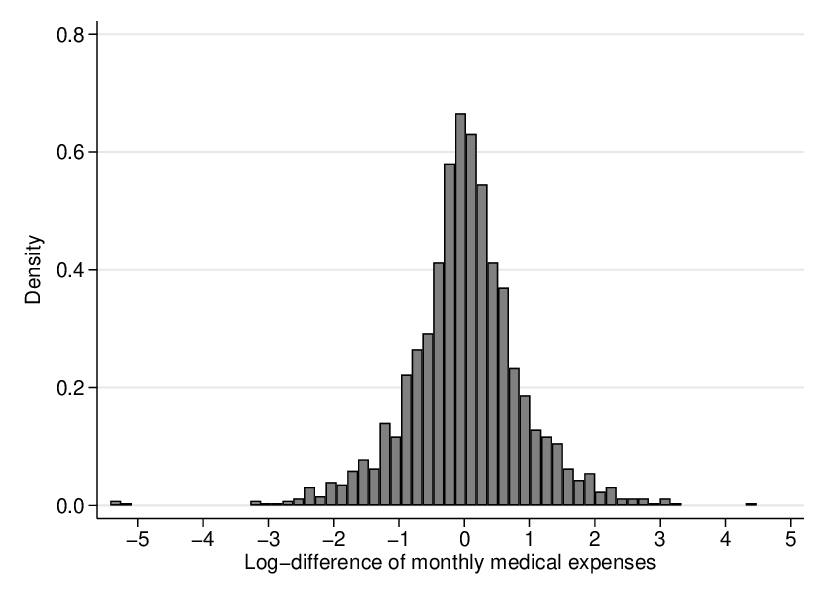}}
  \subfloat[Entertainment]{\label{fig:histentertainment}\includegraphics[width=0.2\textwidth]{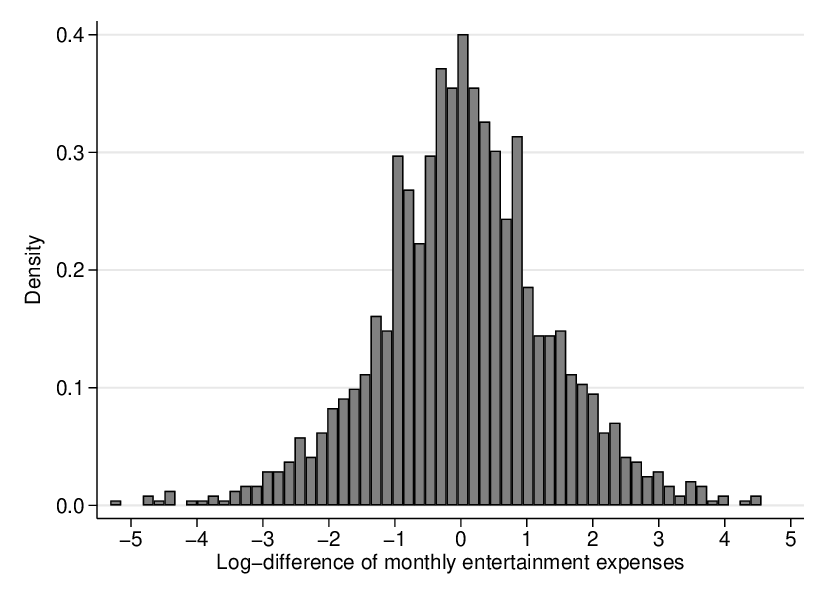}}
  \subfloat[Transportation]{\label{fig:histtransportation}\includegraphics[width=0.2\textwidth]{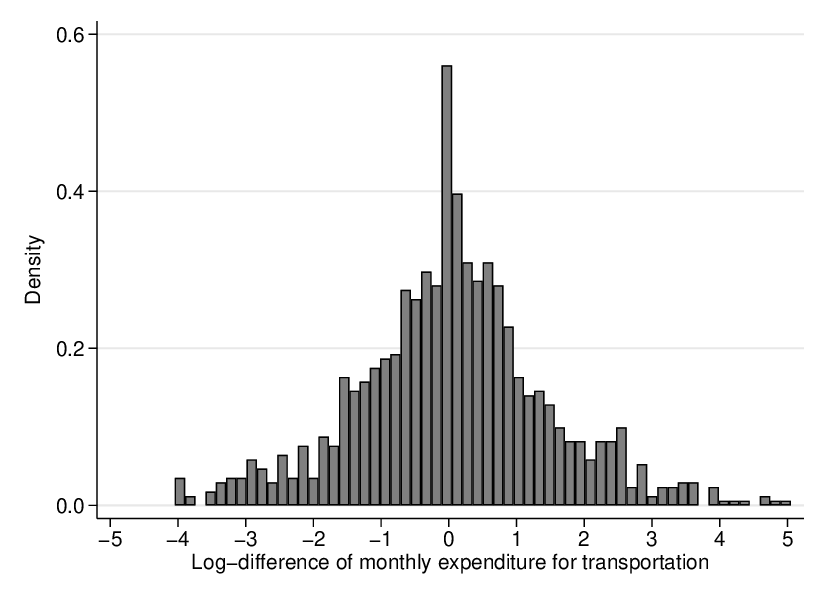}}
  \subfloat[Miscellaneous]{\label{fig:histother}\includegraphics[width=0.2\textwidth]{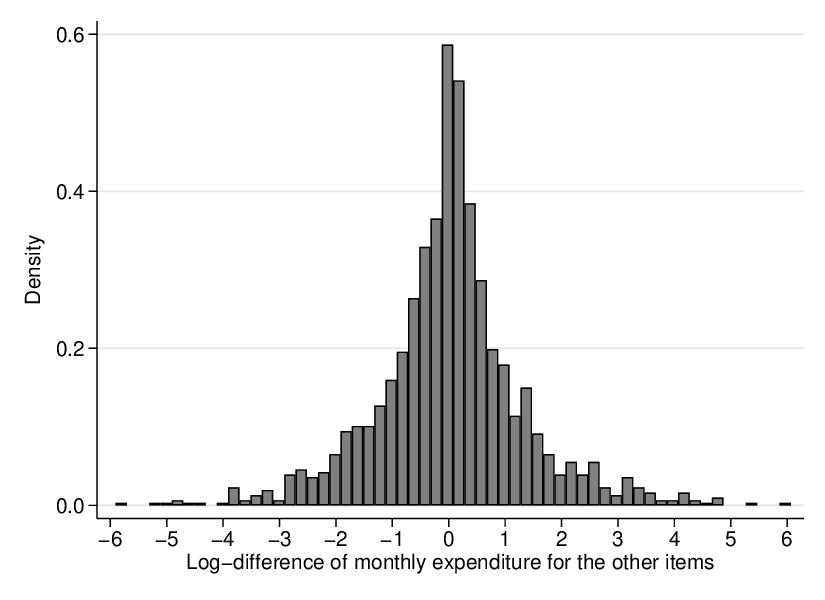}}
\caption{Distribution of the log-difference of the 10 subcategories}
\label{fig:ldsub}
\scriptsize{\begin{minipage}{450pt}
Notes:
The distribution of the log-differences in monthly expenditure for the 10 subcategories listed in panel A of Table~\ref{tab:sum} is shown in the figures.
Sources: Created by the author from the Municipal Bureau of Labor Research of Osaka (1919--1920).
\end{minipage}}
\end{figure}

\begin{figure}[h]
\centering
\captionsetup{justification=centering}
  \subfloat[Withdrawals and gifts]{\label{fig:inc_other}\includegraphics[width=0.25\textwidth]{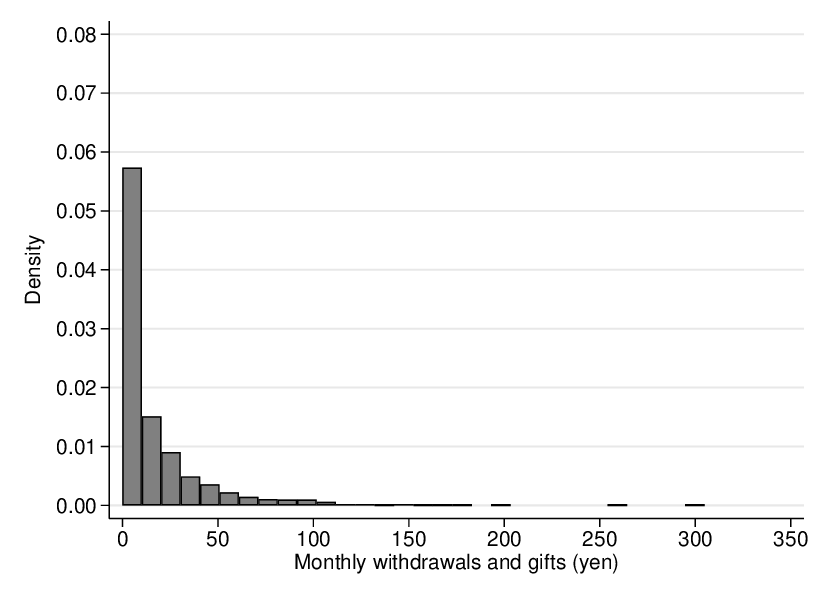}}
  \subfloat[Deposits to savings]{\label{fig:exp_saving}\includegraphics[width=0.25\textwidth]{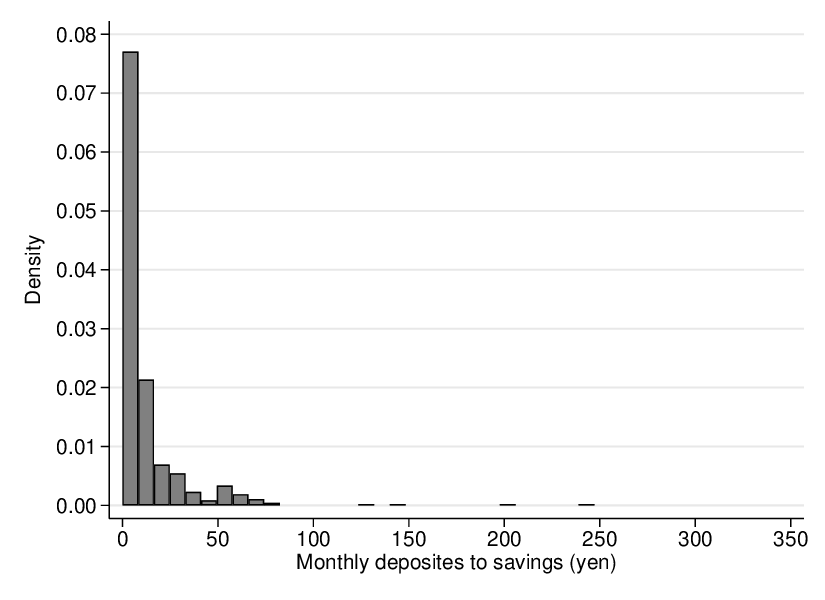}}
  \subfloat[Borrowing]{\label{fig:inc_loan}\includegraphics[width=0.25\textwidth]{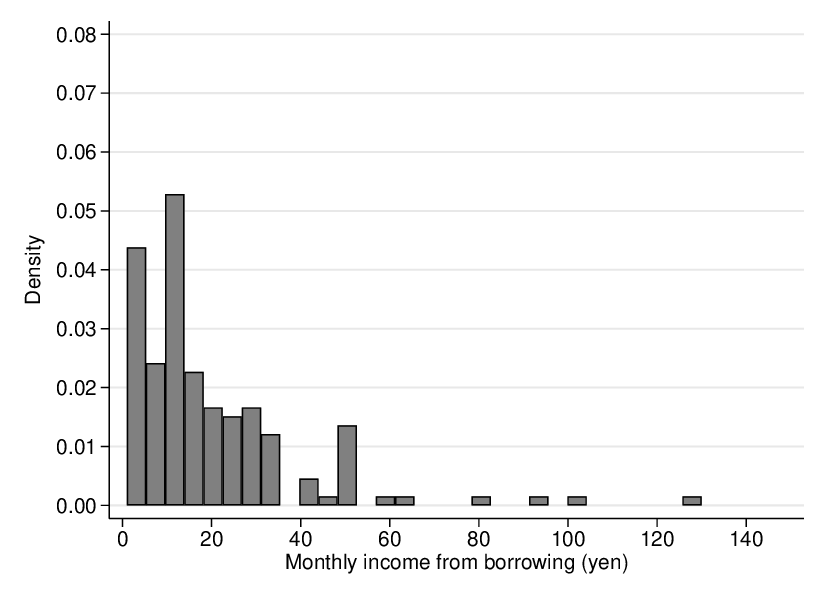}}
  \subfloat[Liquidation of loans]{\label{fig:exp_liquidation}\includegraphics[width=0.25\textwidth]{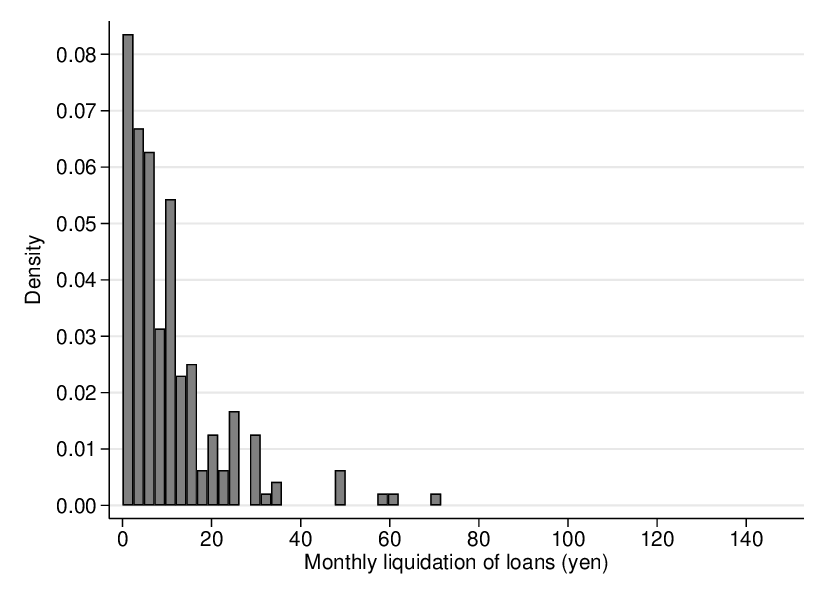}}
\caption{Distribution of monthly income from temporary sources\\ and the corresponding expenditure (yen)}
\label{fig:inc_sources}
\scriptsize{\begin{minipage}{450pt}
Notes:
Figure~\ref{fig:inc_loan} illustrates the distribution of monthly withdrawals and gifts.
Figure~\ref{fig:exp_saving} shows the distribution of monthly deposits to savings.
Figure~\ref{fig:inc_loan} illustrates the distribution of monthly temporary income from borrowing.
Figure~\ref{fig:exp_liquidation} shows the distribution of monthly liquidation of loans.
The range of the x-axis is fixed at zero to 350 (150) in the savings (borrowing) category.
Sources: Created by the author from the Municipal Bureau of Labor Research of Osaka (1919--1920).
\end{minipage}}
\end{figure}

Figures~\ref{fig:dinc} and \ref{fig:exp} show the monthly disposable income and expenditure of RLR households, respectively.
The distribution of both figures is right skewed, thus showing the typical distribution of income and expenditure.
The mean values of disposable income and expenditure are $91.3$ and $88.7$ yen, respectively.
Since outliers are excluded as noted in the main text, no specific observations take extremely high values. 
Figures~\ref{fig:lnddinc} and \ref{fig:lndexp} show the distributions of the log-differences in disposable income and expenditure, respectively.
Figure~\ref{fig:scat_exp} describes the correlation between the log-differences in monthly disposable income and expenditure.
Figure~\ref{fig:ldsub} presents the log-difference in monthly expenditure for the 10 subcategories.
Figure~\ref{fig:inc_sources} describes the distribution of monthly income from savings withdrawals and gifts (a); deposits to savings (b); borrowing (c); and liquidation of loans (d), confirming that no systematic outliers exist.


\subsection{Monthly Income of Adult Male Factory Workers}\label{sec:seca2}

\begin{table}[!h]
\def\arraystretch{1.0}
\centering
\begin{center}
\caption{Calculating the Monthly Income of Adult Male Factory Workers}
\label{tab:wage}
\scriptsize
\scalebox{1.0}[1]{
\begin{tabular}{lrrrrrr}
\toprule
\multicolumn{7}{l}{\textbf{Panel A: Components in equation 16}}\\
&
&
&\multicolumn{2}{c}{$\textit{Annual Working Days}_{s}$}
&\multicolumn{2}{c}{$\textit{Bonus}_{s}$ (yen)}\\
\cmidrule(rr){4-5}\cmidrule(rr){6-7}
&(1) $\textit{Wage}^{Adult}_{s}$
&(2) $\textit{Miscellaneous}$
&(3)
&(4)
&(5)
&(6)\\
Sector				& (yen)			& (yen)			&Conservative		&$+12$ days 	&(3) used 	&(4) used\\

\hline
Textile				&1.83			&0.15			&322		&334		&49.1		&50.9\\
Machine				&2.16			&0.31			&320		&332		&57.6		&59.8\\
Chemical			&2.01			&0.20			&331		&343		&55.4		&57.5\\
Food				&2.02			&0.13			&322		&334		&54.2		&56.2\\
Miscellaneous		&2.08			&0.20			&328		&340		&56.9		&58.9\\\hline
&&&&&&\\
\multicolumn{7}{l}{\textbf{Panel B: Monthly income of adult male factory workers}}\\
&\multicolumn{4}{c}{$\textit{Monthly Income}_{s}$ (yen)}
&\multicolumn{2}{c}{}\\
\cmidrule(rr){2-5}
			&\multicolumn{2}{c}{(1)}				&\multicolumn{2}{c}{(2)}	&\multicolumn{2}{c}{}\\
Sector		&\multicolumn{2}{c}{$\textit{Annual Working Days}_{s}$ = (3)}&\multicolumn{2}{c}{$\textit{Annual Working Days}_{s}$ = (4)}&\multicolumn{2}{c}{}\\
\cmidrule(lrrrr){1-5}
Textile				&\multicolumn{2}{c}{57.2}		&\multicolumn{2}{c}{59.4}	&\multicolumn{2}{c}{}\\
Machine				&\multicolumn{2}{c}{70.7}		&\multicolumn{2}{c}{73.3}	&\multicolumn{2}{c}{}\\
Chemical			&\multicolumn{2}{c}{65.6}		&\multicolumn{2}{c}{68.0}	&\multicolumn{2}{c}{}\\
Food				&\multicolumn{2}{c}{62.2}		&\multicolumn{2}{c}{64.5}	&\multicolumn{2}{c}{}\\
Miscellaneous		&\multicolumn{2}{c}{67.1}		&\multicolumn{2}{c}{69.5}	&\multicolumn{2}{c}{}\\\bottomrule
\end{tabular}
}
{\scriptsize
\begin{minipage}{410pt}
Notes:
$\textit{Wage}^{Adult}_{s}$ in column (1) in panel A is calculated using equation~\ref{eqn:wage2}.
$\textit{Miscellaneous}$ in column (2) in panel A is the average daily income from sources other than daily wages (see \textit{Miscellaneous Income} section in Online Appendix~\ref{sec:seca2}).
$\textit{Annual Working Days}_{s}$ in columns (3) and (4) in panel A indicate the annual average number of working days.
$\textit{Bonus}_{s}$ (yen) in columns (5) and (6) in panel A are the one-month equivalent bonus calculated using the average number of working days listed in columns (3) and (4), respectively.
$\textit{Monthly Income}_{s}$ in columns (1) and (2) in panel B are calculated based on equation~\ref{eqn:wage1}.\\
Sources:
Data used to calculate the wage are from Osaka City (Osaka City Office 1921 pp.~8(44)--8(45); 1922, pp.~8(46)--8(47)).
Data on miscellaneous income are from the Department of Social Affairs, Osaka City Office (1923, pp.~143--156). Data on working days are from the Department of Social Affairs, Osaka City Office (1923, p.~196).
\end{minipage}
}
\end{center}
\end{table}

\subsubsection*{\textit{Prediction Equation}}

The monthly income of adult male factory workers in the manufacturing sector ($s$), listed in column (2) of panel B in Table~\ref{tab:tab1}, is calculated as follows:
\begin{equation}\label{eqn:wage1}
\textit{Monthly Income}_{s} = \frac{(\textit{Wage}^{Adult}_{s} + \textit{Miscellaneous})\times \textit{Annual Working Days}_{s} + \textit{Bonus}_{s}}{12},
\end{equation}
where $\textit{Wage}^{Adult}$ is the average daily wage of adult male factory workers, $\textit{Miscellaneous}$ indicates the average daily miscellaneous income other than wage, $\textit{Annual Working Days}$ is the average number of annual working days, and $\textit{Bonus}$ is the bonus.
The calculation method for each component is summarized in the subsections below.

\subsubsection*{\textit{Average Daily Wage of Adult Male Factory Workers}}

The average daily wages of adult male factory workers in the manufacturing sectors are not explicitly reported in the manufacturing census.
Therefore, following the method suggested by Ohkawa et al. (1967), I systematically back calculated it as follows:
\begin{equation}\label{eqn:wage2}
\textit{Wage}^{Adult}_{s} = \frac{\textit{Wage}^{Average}_{s} (\textit{Workers}^{Adult}_{s} + \textit{Workers}^{Child}_{s}) - \textit{Wage}^{Child}_{s} \times \textit{Workers}^{Child}_{s}}{\textit{Workers}^{Adult}_{s}},
\end{equation}
where $\textit{Wage}^{Average}$ is the average daily wage of male factory workers, $\textit{Wage}^{Child}$ is the average minimum daily wage, and $\textit{Workers}^{Adult}$ and $\textit{Workers}^{Child}$ are the number of male factory workers aged 20 years and older (\textit{seinen k\=o}) and less than 20 years (\textit{sh\=onen k\=o}), respectively.
The minimum wage is used for child workers because they had received bottom wages in factories (Department of Social Affairs, Osaka City Office 1923, p.~159).
All these available data are taken from the manufacturing census reported by Osaka City Office (1921
pp.~8(44)--8(45); 1922, pp.~8(46)--8(47)).
The RLR periods range from 1919 to 1920 (Section~\ref{sec:sec31}).
Thus, I calculated the $\textit{Wage}^{Average}$ for both 1919 and 1920 and then weighted both figures using analytical weights based on the number of observations in the RLR sample, say four (for 1919) to five (for 1920), to obtain the average daily wage of male factory workers.\footnote{The results are materially similar if I use equal weights (i.e., one (for 1919) to one (for 1920)) instead of the analytical weight.}
Column (1) in panel A of Table~\ref{tab:wage} lists the calculated wage in each manufacturing sector.
These figures are reasonably higher than the average daily wage of male factory workers in each sector in Osaka city (Osaka City 1921 pp.~8(44)--8(45); 1922, pp.~8(46)--8(47)).

\subsubsection*{\textit{Miscellaneous Income}}

In addition to daily wages, factory workers received a small amount of miscellaneous income from their factories, which includes allowances, division of profits (\textit{rijyun bunpai}), and payments for overtime hours worked.\footnote{Although the division of profits is not very common at that time, most of the factories had payed any of those miscellaneous incomes to the workers (Department of Social Affairs, Osaka City Office 1923, pp.~138--139).}
Although the systematic statistics on these sources by manufacturing sector are scarce, an official factory survey documented the average daily miscellaneous income for 156 factories in Osaka city (Department of Social Affairs, Osaka City Office 1923, pp.~143--156).
While this survey covers all factories with more than 100 workers, the data obtained from this survey should not be critically biased upward because the average number of workers per factory was roughly 156 at that time, which is sufficiently above the threshold value (Department of Social Affairs, Osaka City Office 1926, p.~1).\footnote{If the provision and amount of miscellaneous income depended on the scale of the factories, then the figures obtained from this survey may be biased upward. However, there is no statistically significant difference in the average daily wages of adult males (excluding miscellaneous income) between the factories that provided miscellaneous income and those that did not.~In addition, there is no clear positive correlation between the average daily wages of adult males (excluding miscellaneous income) and the average miscellaneous income. These results suggest that the potential upward bias is not remarkable.}
Despite this, I choose the 25th percentile as the reference value of the miscellaneous income rather than the mean or median to provide conservative estimates.\footnote{Note that the slight change in this value does not disturb the overall interpretation because the income from these sources was substantially smaller than the daily wage; for instance, a $\pm 0.05$ yen (i.e., 25\%) change in this term results in only a few percentage change in the calculated $\textit{Monthly Income}_{s}$.}
As shown in column (2) of Table~\ref{tab:wage}, these figures range from roughly $0.1$ to $0.3$, which are indeed similar to those documented in the other available sources (Department of Social Affairs, Osaka City Office 1922, p.~74-76), which supports plausibility of this estimate.

\subsubsection*{\textit{Annual Working Days}}

The average number of annual working days by manufacturing sector listed in column (3) of panel A of Table~\ref{tab:wage} is taken from an official report on the factory survey in Osaka city that I used in the prior subsection (Department of Social Affairs, Osaka City Office 1923, pp.196-197).
The figures in column (3) are conservative in the following ways.
First, the average number of days off reported in the manufacturing census for Osaka was approximately 2.4 days per month, which is one day less than the average obtained herein (Secretariat of Agriculture and Commerce 1921, pp.~156--157).
Second, some factories even paid wages for these holidays (Department of Social Affairs, Osaka City Office 1923, pp.198--199).
Therefore, while I use the conservative value listed in column (3) for the main calculation, I also consider an alternative case in column (4), which adds 12 days (i.e., one day per month) to those figures.

\subsubsection*{\textit{Bonus}}

The total amount of bonus for factory workers in Osaka city around 1920 was equivalent to one month's pay (Tada 1991a, pp.40--49), which was usually paid in December (Tada 1991b, p.9; Department of Social Affairs, Osaka City Office 1922, p.197).
Columns (5) and (6) in panel A of Table~\ref{tab:wage} show the one-month equivalent bonus under the conservative setting using the annual working days listed in column (3) and the one-month equivalent bonus calculated using column (4), respectively.

\subsubsection*{\textit{Monthly Income}}

Column (1) in panel B of Table~\ref{tab:wage} shows the calculated $\textit{Monthly Income}_{s}$, which is also reported in the main text (column (2) of panel B in Table~\ref{tab:tab1}).
Column (2) in the same panel lists the calculated figures under the tolerant setting in terms of the number of annual working days.

\subsection{Comparing the RLR Sample with the Census and Survey Samples}\label{sec:seca3}

\def\arraystretch{1.0}
\begin{table}[h!]
\begin{center}
\captionsetup{justification=centering}
\caption{Household Size, Monthly Income, and Expenditure in the Manufacturing Sector across Regions: Comparing the RLR Sample with the Census and Survey Samples}\label{tab:taba2}
\scriptsize
\scalebox{0.93}[1]{
\begin{tabular}{lrrrr}\toprule
\multicolumn{5}{l}{\textbf{Panel A: Population census}}\\
&\multicolumn{2}{r}{Household size (people) in}	&&\\
&\multicolumn{2}{r}{manufacturing industry}		&\multicolumn{2}{r}{Number of	households}\\\hline
RLR sample										&&4.00			&&237	\\
												&&[3.79, 4.21]	&&		\\
Population census								&&&&\\
\hspace{10pt}Representative large cities			&&&&\\
\hspace{20pt}Osaka							&&3.99	&&107,340	\\
\hspace{20pt}Tokyo							&&4.19	&&168,226	\\
\hspace{20pt}By wider region					&&&&\\
\hspace{30pt}Midwest (Osaka, Kobe, Kyoto)		&&4.01	&&208,437	\\
\hspace{30pt}Mideast (Tokyo, Yokohama)		&&4.18	&&198,357	\\
&&&&\\
\hspace{10pt}Smaller cities in other regions		&&&&\\
\hspace{20pt}Midland (Nagoya)					&&4.31	&&36,286	\\
\hspace{20pt}Southwest (four cities)			&&4.22	&&45,960\\
\hspace{20pt}Northeast	 (four cities)				&&4.63	&&22,546\\\hline
&&&&\\
\multicolumn{5}{l}{\textbf{Panel B: Survey of the Living Conditions of Factory Workers}}\\
\multicolumn{5}{l}{Survey area: Eight cities located from the southwest to the mideast regions}\\
\multicolumn{5}{l}{Survey subject: Factory worker households with 4--6 people; head earned more than $30$ yen per month}\\
\multicolumn{5}{l}{Survey month and year: February and March 1921}\\
&\multicolumn{3}{c}{Mean values}&\\
\cmidrule(rrr){2-4}
								&(1) Monthly household	&(2) Monthly household	&(3) Household size	&Number of\\
								& Income (yen)			&expenditure (yen)		& (people)			&households\\\hline
RLR sample						&103.17				&91.72					&4.78				&82\\
								&[96.57, 109.78]		&[86.45, 96.99]		&[4.66, 4.91]		&\\
LCFW sample					&&&&\\
\hspace{10pt}Representative large cities			&&&&\\
\hspace{20pt}Midwest (Osaka, Kobe, Kyoto)		&103.18	&90.45		&4.84	&603\\
\hspace{20pt}Mideast (Tokyo, Yokohama)		&103.62	&88.97		&4.72	&399\\
&&&&\\
\hspace{10pt}Smaller cities in other regions	&&&&\\
\hspace{20pt}Midland (Nagoya)					&104.40	&82.02		&4.63	&155\\
\hspace{20pt}Southwest (Nagasaki, Fukuoka)	&101.04	&87.37		&5.08	&256\\\bottomrule
\end{tabular}
}
{\scriptsize
\begin{minipage}{446pt}
Notes:\\
1. Panel A compares the mean household size in the RLR sample with the mean size of households in the manufacturing sector in the large cities available from the 1920 population census.
In the census, large cities are defined as cities with populations greater than $100,000$.
The southwest region includes Hiroshima, Kure, Nagasaki, and Kagoshima.
The northeast region includes Hakodate, Otaru, Sapporo, and Sendai.
The number of households obtained from the census indicates the total number of households in the manufacturing sector.\\
2. Panel B shows the mean monthly income, expenditure, and household size calculated from the LCFW samples.\\
Altogether, 82 RLR households satisfying the sampling criteria in the LCFW (i.e., household with 4--6 people and the head earned more than 30 yen per month) are used.
The monthly income and expenditure in the RLR sample are calculated using the data from February and March 1920 to match the months of the LCFW.
The mean monthly income and expenditure from the survey are deflated using the consumer price index provided by the Bank of Japan.\\
3. The figures in brackets are 95\% confidence intervals.\\
Sources:
Calculated by the author from the Municipal Bureau of Labor Research of Osaka (1919--1920); Statistics Bureau of the Cabinet (1929b, pp. 320--385) (for Panel A); Bureau of Social Welfare (1923) (for Panel B).
Data on the consumer price index are obtained from Bank of Japan (1986).
\end{minipage}
}
\end{center}
\end{table}

Panel A of Table~\ref{tab:taba2} provides the mean size of households with heads working in the manufacturing industry across the Japanese archipelago.
The mean values of Osaka and Tokyo are $3.99$ and $4.19$, respectively, close to those of the RLR sample ($4.00$ with a 95\%CI of $[3.79, 4.21]$).
At the regional level, the mean values for the Midwest region (including Osaka, Kobe, and Kyoto) and Mideast region (including Tokyo and Yokohama) show similar values of $4.01$ and $4.18$, respectively.
This makes sense because the manufacturing industries in these cities were at a similar developmental stage.
In fact, smaller and less developed cities in other regions had larger households.
For example, Nagoya, a medium-sized city in the midland region, and the other smaller provincial cities in the southwest and northeast regions have larger households ($4.31$, $4.22$, and $4.63$, respectively).
Thus, applying the findings from the RLR sample to these provincial cities could be misleading because factory worker households in both regions might have had different preferences for household consumption compared to the smaller households in populated cities.

In panel B of Table~\ref{tab:taba2}, I provide additional evidence on the representativeness of the RLR sample by using the official report of a large household survey -- the Survey of Living Conditions of Factory Workers (LCFW), conducted by the Bureau of Social Welfare in 1921.
The LCFW surveyed factory worker households with families consisting of 4--6 people and with heads who earned more than 30 yen per month.
Focusing on households with families is useful because the proportion of single workers in the manufacturing sector was considerably low: 1--2\% both in the population census and RLR sample (Statistics Bureau of the Cabinet 1929b, pp.~320--325).\footnote{The comparison for the households with families with 2--3 people, which is not included in the LCFW, should be similar to that in panel B of Table~\ref{tab:tab1}. Note that the average monthly earning of households with a couple (or couple and a small child) is similar to that of the heads in breadwinner households  (Tada 1991a).}
Moreover, the phenomenon of a head earning less than 30 yen per month was rare, as they were usually considered as poor and belonging to the bottom 1\% (Osaka Prefecture 1931, p.~5): in fact, these households account for $0.8$\% of the RLR sample.
Another advantage of the LCFW is that it surveys all representative large cities as well as some smaller cities, thereby allowing me to make a comparison in the same way as in panel A of Table~\ref{tab:taba2}.\footnote{The comprehensiveness of the LCFW could be because it was a systematically designed survey that offered reliable reference materials for planning the initial Health Insurance Act (Tada 1991b, p.~7).}
To ensure a precise comparison, I trim the RLR sample to $82$ households, satisfying the sampling criteria of the LCFW.
In addition, I limit it to February and March to match the survey months of the LCFW.

Columns (1)--(3) in panel B of Table~\ref{tab:taba2} present the mean values of monthly household income, expenditure, and household size, showing evidence that the RLR sample can approximate the mean values of household income, expenditure, and household size in representative large cities in the midwest and mideast regions, as confirmed in panel A of Table~\ref{tab:taba2}.
Although some figures in smaller cities in the midland and southwestern regions show values that are similar to the large cities, monthly expenditure in Nagoya ($82.02$ yen) and household size in the southwestern region ($5.08$) deviate significantly from the RLR samples. This implies that the households in both regions showed different consumption behaviors compared to those in the representative large cities (Section~\ref{sec:sec31}).

As discussed in the main text, the primary target of this study is factory worker households in Osaka city.
Given the similarities, however, it might be conceivable to view factory worker households in other large cities as the secondary target, although with care taken to adapt the findings from the RLR sample.

\section{Empirical Analysis Appendix}\label{sec:secb}
\setcounter{table}{0} \renewcommand{\thetable}{C.\arabic{table}}
\setcounter{figure}{0} \renewcommand{\thefigure}{C.\arabic{figure}}

\subsection{Additional Results: Robustness to Preference Shifts}\label{sec:secb1}

I provide additional evidence on the robustness of my main results to potential preference shifts.
The most influential event inducing preference shifts is the loss of the head, which can cause both a negative idiosyncratic income shock and a reduction in food consumption.
I check the frequency of household head deaths during the sample period in the following two ways.

First, I utilize the data on the household size.
In the RLR documents, there are two types of reported household sizes: raw household size is reported in the initial survey month, and adult equivalent household size is reported every month.
If the head died, then the adult equivalent size should decrease accordingly.
However, I confirmed that the adult equivalent household size remained unchanged throughout the survey period in all the households. This suggested that no heads had died during this period.\footnote{If a head died and an adult man became the new head in the same month, the adult equivalent household size is unchanged. However, this replacement of a head is not likely because the head's occupation rarely changed during the sample period, as explained previously. Moreover, even if such a replacement occurred, there should be no preference shift because the family size was stable regardless of the loss of the head.}
This is consistent with the fact that most of the factory workers in Osaka city were in their 20s--30s, and the average life expectancy of males at age 20 was approximately 40 years at that time.\footnote{See Department of Social Affairs, Osaka City Office (1923, p.~19) and Ministry of Health, Labour and Welfare, (\textit{Life Table}, database) for the statistics.}

Second, I use the information on heads' occupation to check whether there were any losses of heads.
I undertook this exercise because the adult equivalent household size reported every month was possibly a repetition of the raw household size reported in the initial period.
If a head died, then the head's occupation would change because the wife would have become the breadwinner.\footnote{
Note that the wives were rarely employed in the manufacturing sector (Tada 1991a).}
I found that there were only six heads who changed their occupation (industry) during the sampled periods.
However, these changes were not permanent but one-month temporary changes.
This means that these changes were not caused by the losses of heads.

\def\arraystretch{1.0}
\begin{table}[h]
\begin{center}
\captionsetup{justification=centering}
\caption{Results of Estimating Income Elasticities:\\ Alternative Specification Including Additional Control Variable}
\label{tab:mace_rob_ac}

\footnotesize
\scalebox{1.0}[1]{
\begin{tabular}{lrlc}\toprule
&\multicolumn{2}{c}{Disposable income}&\\
\cmidrule(rl){2-3}
\textbf{Result from the alternative specification}&Coef.&Std. error&Observations\\ \hline
\hspace{10pt}Total consumption 						&0.392	&[0.037]***	&1880  \\
&&&\\
\hspace{10pt}Food										&0.139	&[0.032]***	&1880  \\
&&&\\
\hspace{10pt}Housing									&0.074	&[0.043]*		&1851  \\
&&&\\
\hspace{10pt}Utilities									&0.293	&[0.106]***	&1733  \\
&&&\\
\hspace{10pt}Furniture									&0.663 &[0.165]***	&1560  \\
&&&\\
\hspace{10pt}Clothes									&0.679	&[0.113]***	&1840  \\
&&&\\
\hspace{10pt}Education									&0.091	&[0.129]		&789  \\
&&&\\
\hspace{10pt}Medical expenses							&0.378	&[0.077]***	&1866  \\
&&&\\
\hspace{10pt}Entertainment expenses					&0.552	&[0.096]***	&1809  \\
&&&\\
\hspace{10pt}Transportation							&0.315	&[0.129]**		&1512  \\
&&&\\
\hspace{10pt}Miscellaneous								&0.524	&[0.135]***	&1854  \\\bottomrule
\end{tabular}
}
{\scriptsize
\begin{minipage}{375pt}
***, **, and * denote statistical significance at the 1\%, 5\%, and 10\% levels, respectively.
Standard errors in brackets are clustered at the household level.\\
Notes: 
This table shows the results for the alternative specification of equation~\ref{eqn:eq2}: the regressions of the 11 measures of log-transformed consumption on log-transformed disposable income and the family size controls, occupational change dummy, household fixed effects, and month-year fixed effects.
The occupational change dummy is an indicator variable that takes the value of one for the year-month cells of six households whose heads had temporarily changed their occupations.
The family size controls are interacted with the quarter dummies.
The estimated coefficients on log-transformed disposable income are listed in the second column (Coef.).
\end{minipage}
}
\end{center}
\end{table}

Next, I quantitatively test the potential influence of preference shifts due to changes in the household size in two ways.
First, I include an indicator variable for the six year-month cells of six households whose heads had temporary changed their occupations, in equation~\ref{eqn:eq2}.
The results are shown in Table~\ref{tab:mace_rob_ac}.
The estimates are virtually identical to those listed in Table~\ref{tab:mace}.\footnote{I have also confirmed that the results are unchanged from my main results if I simply excluded these six households.}
Second, I exclude all the family size controls from equation~\ref{eqn:eq2}.
If the preference shifts due to changes in family size are the cause of the endogeneity in the regressions, my main results presented in Table~\ref{tab:mace} should substantially change after excluding the family size controls.
Table~\ref{tab:mace_rob_exc} shows the results from the specification excluding all these controls.
Again, the estimates are materially similar to those listed in Table~\ref{tab:mace}.

\def\arraystretch{1.0}
\begin{table}[h]
\begin{center}
\captionsetup{justification=centering}
\caption{Results of Estimating Income Elasticities:\\ Alternative Specification Excluding Family Size Controls}
\label{tab:mace_rob_exc}

\footnotesize
\scalebox{1.0}[1]{
\begin{tabular}{lrlc}\toprule
&\multicolumn{2}{c}{Disposable income}&\\
\cmidrule(rl){2-3}
\textbf{Result from the alternative specification}&Coef.&Std. error&Observations\\ \hline
\hspace{10pt}Total consumption 						&0.394	&[0.038]***	&1880  \\
&&&\\
\hspace{10pt}Food										&0.140	&[0.032]***	&1880  \\
&&&\\
\hspace{10pt}Housing									&0.086	&[0.047]*		&1851  \\
&&&\\
\hspace{10pt}Utilities									&0.283	&[0.106]***	&1733  \\
&&&\\
\hspace{10pt}Furniture									&0.645 &[0.165]***	&1560  \\
&&&\\
\hspace{10pt}Clothes									&0.684	&[0.111]***	&1840  \\
&&&\\
\hspace{10pt}Education									&0.113	&[0.124]		&789  \\
&&&\\
\hspace{10pt}Medical expenses							&0.373	&[0.075]***	&1866  \\
&&&\\
\hspace{10pt}Entertainment expenses					&0.553	&[0.094]***	&1809  \\
&&&\\
\hspace{10pt}Transportation							&0.306	&[0.128]**		&1512  \\
&&&\\
\hspace{10pt}Miscellaneous								&0.529	&[0.134]***	&1854  \\\bottomrule
\end{tabular}
}
{\scriptsize
\begin{minipage}{374pt}
***, **, and * denote statistical significance at the 1\%, 5\%, and 10\% levels, respectively.
Standard errors in brackets are clustered at the household level.\\
Notes: 
This table shows the results for the alternative specification of equation~\ref{eqn:eq2}: the regressions of the 11 measures of log-transformed consumption on log-transformed disposable income as well as on the household fixed effects and month-year fixed effects.
The estimated coefficients on log-transformed disposable income are listed in the second column (Coef.).
\end{minipage}
}
\end{center}
\end{table}

The foregoing results support the evidence that the family size in my RLR sample had been relatively stable and thus, the preference shifts should not disturb the main findings in this paper.
Despite this, the estimates listed in Table~\ref{tab:mace_rob_exc} are slightly larger than those listed in Table~\ref{tab:mace}, albeit these differences are negligible.
Therefore, to be conservative, I prefer to include the family size controls in equations~\ref{eqn:eq2} (Section~\ref{sec:sec41}).

\subsection{Additional Results: Alternative Definition of the Shock Variable}\label{sec:secb2}

My baseline specification uses disposable income as the shock variable (Section~\ref{sec:sec41}). 
In this subsection, I use an indicator variable for a negative income shock instead of disposable income in equation~\ref{eqn:eq2} to test whether my main results are derived from negative rather than positive shocks.
Specifically, the indicator variable takes one if disposable income is below households' mean disposable income, whereas it takes zero if disposable income is equal to or above the mean.
Intuitively, this indicator switches to one if the household experiences a negative idiosyncratic shock on its income relative to its potential permanent income.
One can expect that consumption changes little with positive shocks, but co-moves with income following negative shocks, as in Deaton (1991).
Therefore, the estimated coefficient on this indicator variable should be negative, which means that consumption responds to negative shocks relative to positive shocks (and the case of no change in income).

Table~\ref{tab:mace_rob} presents the results, which confirm that the estimates are negative in most cases.
The estimated coefficients across the categories are similar to those of the main results (Table~\ref{tab:mace}).
In Table~\ref{tab:mechanism_rob}, I also use an indicator variable for negative income shocks instead of disposable income in equation~\ref{eqn:eq3}.
The results in this table are also consistent with those in Table~\ref{tab:mechanism}.
Similarly, Table~\ref{tab:mechanism_ls_rob} shows the results from the specification using an indicator variable for negative shocks to the income of the household head instead of that income in the labor supply equations in Table~\ref{tab:mechanism_ls}.

The foregoing results confirm that my results presented in the main text are robust to the variable definition.
However, one must be careful about the fact that these alternative specifications using the indicator shock variable tend to cut off the useful information in the small changes in disposable income because it is simply rounded to one or zero based on the threshold.
This leads to an identification issue because of the smaller variation in the key indicator variable, especially with few observations.
I acknowledge that the regressions for testing the heterogeneous responses (with respect to the borrowing, clothes, and furniture categories) using the small sample size presented in Table~\ref{tab:mechanism_hetero} are no longer computationally practical.
Despite this, the weight of evidence shown in Tables~\ref{tab:mace_rob}, \ref{tab:mechanism_rob}, and \ref{tab:mechanism_ls_rob} suggests that the definition of the key variable does not matter for those regressions.

\def\arraystretch{1.0}
\begin{table}[h]
\begin{center}
\captionsetup{justification=centering}
\caption{Results of Estimating Income Elasticities:\\ Alternative Specification using Alternative Definition of Shock Variable}
\label{tab:mace_rob}

\footnotesize
\scalebox{1.0}[1]{
\begin{tabular}{lrlc}\toprule
&\multicolumn{2}{c}{Negative Shock (=1)}&\\
\cmidrule(rl){2-3}
\textbf{Result from the alternative specification}		&Coef.		&Std. error&Observations\\ \hline
\hspace{10pt}Total consumption 						&-0.149	&[0.015]***	&1880  \\
&&&\\
\hspace{10pt}Food										&-0.045	&[0.045]***	&1880  \\
&&&\\
\hspace{10pt}Housing									&-0.040	&[0.040]*		&1851  \\
&&&\\
\hspace{10pt}Utilities									&-0.110	&[0.110]**		&1733  \\
&&&\\
\hspace{10pt}Furniture									&-0.300	&[0.087]***	&1560  \\
&&&\\
\hspace{10pt}Clothes									&-0.259	&[0.061]***	&1840  \\
&&&\\
\hspace{10pt}Education									& 0.023		&[0.068]		&789  \\
&&&\\
\hspace{10pt}Medical expenses							&-0.137	&[0.037]***	&1866  \\
&&&\\
\hspace{10pt}Entertainment expenses					&-0.208	&[0.047]***	&1809  \\
&&&\\
\hspace{10pt}Transportation							&-0.109	&[0.064]*		&1512  \\
&&&\\
\hspace{10pt}Miscellaneous								&-0.163	&[0.054]***	&1854  \\\bottomrule
\end{tabular}
}
{\scriptsize
\begin{minipage}{382pt}
***, **, and * denote statistical significance at the 1\%, 5\%, and 10\% levels, respectively.
Standard errors in brackets are clustered at the household level.\\
Notes: 
This table shows the results for the alternative specification of equation~\ref{eqn:eq2}: the regressions of the 11 measures of log-transformed consumption on the indicator variable for the negative income shock, family size controls, household fixed effects, and month-year fixed effects.
The family size controls are interacted with the quarter dummies.
The estimated coefficients on log-transformed disposable income are listed in the second column (Coef.).
\end{minipage}
}
\end{center}
\end{table}
\def\arraystretch{1.0}
\begin{table}[h]
\begin{center}
\captionsetup{justification=centering}
\caption{Results of Testing the Risk-Coping Mechanisms:\\ Alternative Specification using Alternative Definition of Shock Variable
\label{tab:mechanism_rob}
}
\footnotesize
\scalebox{0.95}[1]{
\begin{tabular}{lrr}\toprule
\textbf{Panel A: Testing the role of savings}
&\multicolumn{2}{c}{Dependent variable}\\
\cmidrule(rr){2-3}
&\multicolumn{1}{c}{(1) Withdrawals and gifts}&\multicolumn{1}{c}{(2) Deposits to savings}\\\hline
Negative shock (=1)							&17.819***	&-15.195**		\\
											&[5.083]	&[7.729]		\\
Observations								&1,711		&1,711			\\\hline
&&\\
\textbf{Panel B: Testing the role of borrowing}
&\multicolumn{2}{c}{Dependent variable}\\
\cmidrule(rr){2-3}
&\multicolumn{1}{c}{(1) Borrowing}&\multicolumn{1}{c}{(2) Liquidation of}\\
&\multicolumn{1}{c}{}&\multicolumn{1}{c}{loans}\\\hline
Negative shock (=1)								&12.132*	&-7.820*	\\
												&[6.318]	&[4.085]	\\
Observations									&599		&599		\\\hline
&&\\
\textbf{Panel C: Testing the role of pawnshops}
&\multicolumn{2}{c}{Dependent variable}\\
\cmidrule(rr){2-3}
&\multicolumn{1}{c}{(1) Expenditure on}&\multicolumn{1}{c}{(2) Expenditure on}\\
&\multicolumn{1}{c}{clothes}&\multicolumn{1}{c}{furniture}\\\hline
Negative shock (=1)								&-3.323***	&0.180		\\
												&[1.205]	&[1.846]	\\
Observations									&599		&599		\\\bottomrule
\end{tabular}
}
{\scriptsize
\begin{minipage}{420pt}
***, **, and * denote statistical significance at the 1\%, 5\%, and 10\% levels, respectively.
The results from the fixed-effects Tobit model, as proposed by Honor\'e (1992), are reported in columns (1) and (2) in each panel. 
A quadratic loss function is applied for the estimation to ensure computational tractability.
Robust standard errors are in brackets.
Standard errors are clustered at the household level in the linear models.\\
Notes:
Panel A presents the results for the savings category: withdrawals and gifts (column (1)) and deposits to savings (column (2)).
Panel B presents the results for the borrowing category: borrowing (column (1)) and liquidation of loans (column (2)).
Panel C presents the results for expenditure on clothes (column (1)) and furniture (column (2)).
All the regressions in each panel include the indicator variable for the negative income shock, family size controls, household fixed effects, and month-year specific fixed effects.
The family size controls are interacted with the quarter dummies.
\end{minipage}
}
\end{center}
\end{table}
\def\arraystretch{1.0}
\begin{table}[h]
\begin{center}
\captionsetup{justification=centering}
\caption{Results of Testing Labor Supply Adjustments:\\ Alternative Specification using Alternative Definition of Shock Variable
\label{tab:mechanism_ls_rob}
}
\footnotesize
\scalebox{1.0}[1]{
\begin{tabular}{lrrrr}\toprule
\multicolumn{5}{l}{\textbf{Panel A: Testing labor supply adjustments}}\\
&\multicolumn{4}{c}{Dependent variable}\\
\cmidrule(rrrr){2-5}
~~~~~~~~~~~~~~~~~~~~~~~~~~~~~~~~~~~
&\multicolumn{2}{c}{(1) Wife's income}&\multicolumn{2}{c}{(2) Child's income}\\\hline
Negative shock (=1)	&\multicolumn{2}{c}{1.850}	&\multicolumn{2}{c}{9.492***}	\\
					&\multicolumn{2}{c}{[1.361]}	&\multicolumn{2}{c}{[4.291]}	\\
Observations		&\multicolumn{2}{c}{1,627}		&\multicolumn{2}{c}{1,627}		\\\hline
&&&&\\
\multicolumn{5}{l}{\textbf{Panel B: Testing the heterogeneous responses}}\\
&&&&\\
&\multicolumn{4}{c}{Households' mean monthly deposits to savings}\\
\cmidrule(rrrr){2-5}
&\multicolumn{2}{c}{$\leq$ Median}&\multicolumn{2}{c}{$>$ Median}\\
\cmidrule(rr){2-3}\cmidrule(rr){4-5}
&\multicolumn{2}{c}{income from}&\multicolumn{2}{c}{income from}\\
\cmidrule(r){2-2}\cmidrule(r){3-3}\cmidrule(r){4-4}\cmidrule(r){5-5}
&\multicolumn{1}{c}{(1) wife}&\multicolumn{1}{c}{(2) child}&\multicolumn{1}{c}{(3) wife}&\multicolumn{1}{c}{(4) child}\\\hline
Negative shock (=1)	&4.636***	&13.025**	&0.406		&-0.297	\\
					&[1.786]	&[6.613]	&[1.851]	&[6.546]	\\
Observations		&784		&784		&843		&843		\\\bottomrule
\end{tabular}
}
{\scriptsize
\begin{minipage}{340pt}
***, **, and * denote statistical significance at the 1\%, 5\%, and 10\% levels, respectively.
The results from the fixed-effects Tobit model, as proposed by Honor\'e (1992), are reported. 
A quadratic loss function was applied for the estimation to ensure computational tractability.
Robust standard errors are in parentheses.\\
Notes:
Panel A shows the results for the 194 households with three or more family members.
Panel B stratifies households based on the median of households' mean monthly deposits to savings.
Columns (1) and (2) show the results for the 97 households below the median and columns (3) and (4) present the results for the 97 households above the median.
All the regressions in each panel include the indicator variable for the negative shock on the income of the household head, family size controls, household fixed effects, and month-year fixed effects.
The family size controls are interacted with the quarter dummies.
\end{minipage}
}
\end{center}
\end{table}

\clearpage
\subsection{Additional Results: Robustness to Trimming}\label{sec:secb3}

To deal with censoring in the dependent variables, I used the trimmed subsample to investigate the risk-coping mechanisms (Section~\ref{sec:sec51}). I present the evidence on the validity of this procedure as follows.

First, I show that the main findings remain unchanged if I use the full sample.
Table~\ref{tab:mechanism_rob_full} shows the results.
The results in columns (1) and (3) of panel A in Table~\ref{tab:mechanism_rob_full} are similar to those listed in the same columns of panel A in Table~\ref{tab:mechanism}.
This makes sense because the censoring is not severe in these dependent variables for savings.
Columns (1) and (3) of panel B in Table~\ref{tab:mechanism_rob_full} are attenuated but still show statistically significant results.
This attenuation also makes sense because the degree of censoring is much more severe in these dependent variables for borrowing.

Second, the estimates from the fixed-effects Tobit models, listed in columns (2) and (4) of Table~\ref{tab:mechanism_rob_full}, are identical to those listed in Table~\ref{tab:mechanism}.
This is consistent with the fact that this model is robust to the existence of the completely censored units (households) that do not have any within variations in the dependent variable (Honor\'e 1992).
The estimation becomes unfeasible when the number of units that have within variations in the dependent variable is considerably small relative to the number of completely censored units, especially when the model is complex.
The regression for the liquidation of loans listed in column (4) of panel B in Table~\ref{tab:mechanism_rob_full} is the case, in which the valid estimate is computationally unavailable.
Despite this, the result from the linear model in column (3) confirms that trimming the sample does not upset the finding for liquidation of loans.

Finally, Table~\ref{tab:binary} presents the results for the balancing tests using the family size variables listed in panel C of Table~\ref{tab:sum}.
If the trimmed subsample and the remaining samples share similar characteristics, the family size variables should be uncorrelated with the usage of these temporary incomes.
Column (1) of Table~\ref{tab:binary} shows the result from a Probit model that regresses the binary dependent variable for the households receiving income from withdrawals and gifts on the family size controls.
Similarly, Column (2) shows the result for borrowing.
Clearly, all the estimated coefficients are close to zero and statistically insignificant.
The Wald statistics support the null results, confirming that there are no statistically significant differences in the family characteristics between the subsamples.

\def\arraystretch{1.0}
\begin{table}[]
\begin{center}
\captionsetup{justification=centering}
\caption{Results of Testing the Risk-Coping Mechanisms: Results for Full Sample
\label{tab:mechanism_rob_full}
}
\footnotesize
\scalebox{1.0}[1]{
\begin{tabular}{lrrrr}\toprule
\textbf{Panel A: Testing the role of savings}
&\multicolumn{4}{c}{Dependent variable}\\
\cmidrule(rrrr){2-5}
&\multicolumn{2}{c}{}&\multicolumn{2}{c}{Deposits to}\\
&\multicolumn{2}{c}{Withdrawals and gifts}&\multicolumn{2}{c}{savings}\\
\cmidrule(rr){2-3}\cmidrule(rr){4-5}
&\multicolumn{1}{c}{(1)}&\multicolumn{1}{c}{(2)}&\multicolumn{1}{c}{(3)}&\multicolumn{1}{c}{(4)}\\\hline
Disposable income							&-0.183***	&-0.457***	&0.061***	&0.321***	\\
											&[0.037]	&[0.097]	&[0.022]	&[0.088]	\\
Model										&Linear		&Nonlinear	&Linear		&Nonlinear	\\
Observations								&1,880		&1,880		&1,880		&1,880\\\hline
&&&&\\
\textbf{Panel B: Testing the role of borrowing}
&\multicolumn{4}{c}{Dependent variable}\\
\cmidrule(rrrr){2-5}
&\multicolumn{2}{c}{}&\multicolumn{2}{c}{Liquidation of}\\
&\multicolumn{2}{c}{Borrowing}&\multicolumn{2}{c}{loans}\\
\cmidrule(rr){2-3}\cmidrule(rr){4-5}
&\multicolumn{1}{c}{(1)}&\multicolumn{1}{c}{(2)}&\multicolumn{1}{c}{(3)}&\multicolumn{1}{c}{(4)}\\\hline
Disposable income								&-0.024**	&-0.516***	&0.016*	&\multicolumn{1}{c}{--}	\\
												&[0.012]	&[0.143]	&[0.009]	&\multicolumn{1}{c}{--}	\\
Model											&Linear		&Nonlinear	&Linear		&Nonlinear	\\
Observations									&1,880		&1,880		&1,880		&1,880		\\\bottomrule
\end{tabular}
}
{\scriptsize
\begin{minipage}{417pt}
***, **, and * denote statistical significance at the 1\%, 5\%, and 10\% levels, respectively.
The results from the fixed-effects Tobit model, as proposed by Honor\'e (1992), are reported in columns (2) and (4) in each panel. 
A quadratic loss function is applied for the estimation to ensure computational tractability.
Robust standard errors are in brackets.
Standard errors are clustered at the household level in the linear models.\\
Notes:
Panel A presents the results for the savings category: withdrawals and gifts (columns (1) and (2)) and deposits to savings (columns (3) and (4)).
Panel B presents the results for the borrowing category: borrowing (columns (1) and (2)) and liquidation of loans (columns (3) and (4)).
Valid estimate is computationally unavailable in column (4) due to severe censoring.
All the regressions in each panel include the disposable income, family size controls, household fixed effects, and month-year fixed effects.
The family size controls are interacted with the quarter dummies.
\end{minipage}
}
\end{center}
\end{table}
\def\arraystretch{1.0}
\begin{table}[h]
\begin{center}
\captionsetup{justification=centering}
\caption{Results for the Balancing Tests
\label{tab:binary}
}
\footnotesize
\scalebox{1.0}[1]{
\begin{tabular}{lrr}\toprule
&\multicolumn{2}{c}{DV: Indicator variable for the households}\\
&\multicolumn{2}{c}{receiving any income from either}\\
&\multicolumn{2}{c}{withdrawals and gifts or borrowing}\\
\cmidrule(rr){2-3}
&(1) Withdrawals and gifts&(2) Borrowed\\\hline
Size												&0.134		&-0.032\\
												&[0.092]	&[0.075]\\
Children aged 6--9 (\%)							&-0.006	&0.013\\
												&[0.010]	&[0.010]\\
Children aged 10--12 (\%)						&-0.002	&-0.002\\
												&[0.013]	&[0.011]\\
Children aged 13--16 (\%)						&-0.003	&0.006\\
												&[0.010]	&[0.009]\\
Men aged 17+ (\%)								&0.008		&-0.013\\
												&[0.009]	&[0.008]\\
Women aged 17+ (\%)							&-0.002	&-0.001\\
												&[0.008]	&[0.008]\\
Intercept										&0.398		&-0.145\\
												&[0.798]	&[0.659]\\\hline
Wald $\chi^{2}$ statistics for zero slope ($p$-value)		&2.83	(0.8297)	&8.64 (0.1951)	\\
Maximized Log-likelihood							&-97.68	&-134.95\\
Pseudo $R$-squared								&0.0156	&0.0307\\
Number of households							&237		&237	\\\bottomrule
\end{tabular}
}
{\scriptsize
\begin{minipage}{382pt}
The results from Probit models are reported.
Robust standard errors are in brackets. \\
Notes: 
The dependent variable in column (1) is an indicator variable that takes one if the household has received temporary income from withdrawals and gifts.
The dependent variable in column (2) is an indicator variable that takes one if the household has received temporary income from borrowing.
The proportion of children aged 0--5 years (\%) is used as the reference group.
\end{minipage}
}
\end{center}
\end{table}

\subsection{Additional Results: Net Savings and Net Borrowing}\label{sec:secb4}

Columns (1) and (2) of Table~\ref{tab:mechanism_net} present the results for net savings (withdrawals minus deposits) and net borrowing (borrowing minus liquidation of loans), respectively.
As I explained in Section~\ref{sec:sec51}, the estimate for the net savings is a combination of the estimates listed in columns (1) and (3) in panel A of Table~\ref{tab:mechanism}: $-0.202-0.067=-0.269$.
Similarly, the estimate for net borrowing is a combination of the estimates shown in columns (1) and (3) in panel B of the same table: $-0.122-0.072=-0.194$.

\def\arraystretch{1.0}
\begin{table}[]
\begin{center}
\captionsetup{justification=centering}
\caption{Risk-Coping Mechanism Test Results\\: Net Savings and Net Borrowing
\label{tab:mechanism_net}
}
\footnotesize
\scalebox{1.0}[1]{
\begin{tabular}{lrrrr}\toprule
&\multicolumn{2}{c}{Dependent variable}\\
\cmidrule(rr){2-3}
&\multicolumn{1}{c}{(1) Net savings}&\multicolumn{1}{c}{(2) Net borrowing}\\\hline
Disposable income							&-0.269***	&-0.194***	\\
											&[0.040]	&[0.066]	\\
Model										&Linear		&Linear		\\
Observations								&1,711		&599		\\\bottomrule
\end{tabular}
}
{\scriptsize
\begin{minipage}{260pt}
*** denotes statistical significance at the 1\% level.
Standard errors clustered at the household level are in brackets.\\
Notes:
Column (1) reports the estimate for net savings (withdrawals minus deposits).
Column (2) reports the estimate for net borrowing (borrowing minus liquidation of loans).
All the regressions in each panel include the disposable income, family size controls, household fixed effects, and month-year fixed effects.
The family size controls are interacted with the quarter dummies.
\end{minipage}
}
\end{center}
\end{table}

\subsection{Additional Results: Alternative Cut-off Variable}\label{sec:secb5}

Evidence suggests that the type of housing depended on the wealth of the households (Nakagawa 1985, pp.~116--117).
A representative example is that the disadvantaged households with lower assets lived in the tunnel-type single story row houses called \textit{nagaya}, with cheaper rents (Tokyo City Social Welfare Bureau 1921).
This means that the expenditure on housing can be used as a proxy of the wealth level of households.

Using the data on expenditure on housing for investigating the monthly saving behaviors is also an advantage.
As previously explained, the expenditure on housing has little underlying variations in most of the households (Figure~\ref{fig:histhousing}).
Moreover, it is nearly independent from the monthly income (Figure~\ref{fig:housing}).
This means that the expenditure on housing does not co-move with the short-run (monthly) earning and saving behaviors and thus, is a plausible measure for the cut-off variable.

Table~\ref{tab:mechanism_hetero_rob} shows the results.
Column (1) shows the estimate for households with less than or equal to the median of mean monthly expenditure on housing. Meanwhile, column (2) shows the estimate for the households with more than the median.
The results are similar to those listed in columns (1) and (3) of panel A in Table~\ref{tab:mechanism_hetero}, supporting the validity of the cut-off variable used in the main text.

\def\arraystretch{1.0}
\begin{table}[h]
\begin{center}
\captionsetup{justification=centering}
\caption{Results of Testing the Role of Savings:\\ Alternative Thresholds using Expenditure for Housing
\label{tab:mechanism_hetero_rob}
}
\footnotesize
\scalebox{0.95}[1]{
\begin{tabular}{lrr}\toprule
&\multicolumn{2}{c}{Households' mean monthly expenditure on housing}\\
\cmidrule(rr){2-3}
&\multicolumn{1}{c}{$\leq$ Median}&\multicolumn{1}{c}{$>$ Median}\\
\cmidrule(r){2-2}\cmidrule(r){3-3}
&\multicolumn{1}{c}{(1) Withdrawals and gifts}&\multicolumn{1}{c}{(2) Withdrawals and gifts}\\\hline
Disposable income							&-0.568***	&-0.468***	\\
											&[0.114]	&[0.136]	\\
Observations								&724		&987		\\\bottomrule
\end{tabular}
}
{\scriptsize
\begin{minipage}{321pt}
*** denotes statistical significance at the 1\% level.
The results from the fixed-effects Tobit model, as proposed by Honor\'e (1992), are reported. 
A quadratic loss function is applied for the estimation to ensure computational tractability.
Robust standard errors are in brackets.\\
Notes: 
The analytical sample used in panel A of Table~\ref{tab:mechanism} is stratified into two subsamples based on the median of households' mean monthly expenditure on housing: column (1) for 101 households less than or equal to the median, and column (2) for 101 households more than the median.
All the regressions include the disposable income, family size controls, household fixed effects, and month-year specific fixed effects.
The family size controls are interacted with the quarter dummies.
\end{minipage}
}
\end{center}
\end{table}
\def\arraystretch{1.0}
\begin{table}[h]
\begin{center}
\captionsetup{justification=centering}
\caption{Result of Testing the Role of Savings:\\ Alternative Subsample Stratified using Expenditure for Housing
\label{tab:mechanism_hetero_rob_const}
}
\footnotesize
\scalebox{1.0}[1]{
\begin{tabular}{lr}\toprule
&\multicolumn{1}{c}{Households' mean monthly expenditure on housing}\\
\cmidrule(r){2-2}
&\multicolumn{1}{c}{$<$ 25 percentile}\\
\cmidrule(r){2-2}
&\multicolumn{1}{c}{(1) Withdrawals and gifts}\\\hline
Disposable income							&-0.624***	\\
											&[0.217]	\\
Observations								&323		\\\bottomrule
\end{tabular}
}
{\scriptsize
\begin{minipage}{324pt}
*** denotes statistical significance at the 1\% level.
The result from the fixed-effects Tobit model, as proposed by Honor\'e (1992), is reported. 
A quadratic loss function is applied for the estimation to ensure computational tractability.
Robust standard error is in bracket.\\
Notes: 
The analytical sample includes 62 households below the 25th percentile of households' mean monthly expenditure on housing, and that had not received any income from borrowing.
The regression includes the disposable income, family size controls, household fixed effects, and month-year specific fixed effects.
The family size controls are interacted with the quarter dummies.
\end{minipage}
}
\end{center}
\end{table}

Deaton (1991) suggests that precautionary savings protect consumption against income shocks.
This prediction is based on the theory regarding households under a liquidity constraint.
Thus, it cannot be directly applied to households without liquidity constraints (Deaton 1991, p.~1221).
Moreover, since the data on whether the household faced a liquidity constraint are unavailable, it is difficult to identify which household had faced borrowing constraints.
Despite this, I try to assess whether the result in Table~\ref{tab:mechanism_hetero_rob} is robust when I limit the sample to households that potentially faced borrowing constraints.

To do so, I first kept the households that had never received any income from borrowing because those households presumably include households facing a borrowing constraint.
Then, I used the 25th percentile of the mean monthly expenditure on housing as the threshold value, and excluded all the households above this threshold.
This does not mean that all the households under the threshold faced borrowing constraints.
Instead, it implies that those households would be more likely to face the constraints compared to households above this threshold.
Table~\ref{tab:mechanism_hetero_rob_const} shows the result.
The estimate is still negative and statistically significant.

\newpage
\renewcommand{\refname}{References used in the Appendices}

\renewcommand{\refname}{Statistical Reports used in the Appendices}

\end{document}